%% file: main.tex
\documentclass[journal]{IEEEtran}

\usepackage[T1]{fontenc}
\usepackage[utf8]{inputenc}

\usepackage{amsmath}
\usepackage{amssymb}
\usepackage{amsfonts}

\usepackage{graphicx}
\usepackage{booktabs}
\usepackage{multirow}
\usepackage{array}
\usepackage{float}
\usepackage{listings}
\usepackage{xcolor}

\usepackage{cite}
\usepackage{url}

\usepackage[backref=false,bookmarks=false,nolinks=false,breaklinks=true]{hyperref}
\hypersetup{
  hidelinks,
  colorlinks=false,
  breaklinks=true,
  bookmarksopen=false,
  pdftitle={TriSLA: A Preventive and Closed-Loop SLA-Aware Architecture for Multidomain Decision-Making with Explainable Artificial Intelligence in 5G Networks},
  pdfauthor={Abel J. R. Lisboa, Gustavo Z. Bruno, Cristiano B. Both}
}
\usepackage{orcidlink}

\usepackage{acronym}

\usepackage{colortbl}
\usepackage{tikz}

\definecolor{DarkGray}{HTML}{333333}
\definecolor{LightGray}{HTML}{F2F2F2}

\newcommand{\FullCircle}{\tikz[baseline=-0.5ex]\fill (0,0) circle (0.5ex);}
\newcommand{\HalfCircle}{\tikz[baseline=-0.5ex]{\draw (0,0) circle (0.5ex);\fill (0,0.5ex) arc (90:270:0.5ex) -- cycle;}}
\newcommand{\EmptyCircle}{\tikz[baseline=-0.5ex]\draw (0,0) circle (0.5ex);}

\input{acronyms}

\lstdefinelanguage{json}{
    string=[s]{"}{"},
    stringstyle=\color{blue},
    morestring=[s]{'}{'},
    literate=
     *{:}{{{\color{DarkGray}{:}}}}{1}
      {,}{{{\color{DarkGray}{,}}}}{1}
      {\{}{{{\color{DarkGray}{\{}}}}{1}
      {\}}{{{\color{DarkGray}{\}}}}}{1}
      {[}{{{\color{DarkGray}{[}}}}{1}
      {]}{{{\color{DarkGray}{]}}}}{1}
      {"type"}{{{\color{purple}"type"}}}{6}
      {"properties"}{{{\color{purple}"properties"}}}{12}
      {"tenant_id"}{{{\color{purple}"tenant\_id"}}}{11}
      {"slice_service_type"}{{{\color{purple}"slice\_service\_type"}}}{20}
      {"service_requirements"}{{{\color{purple}"service\_requirements"}}}{22}
      {"latency_ms"}{{{\color{purple}"latency\_ms"}}}{12}
      {"throughput_mbps"}{{{\color{purple}"throughput\_mbps"}}}{17}
      {"availability"}{{{\color{purple}"availability"}}}{14}
      {"device_density"}{{{\color{purple}"device\_density"}}}{16}
      {"semantic_context"}{{{\color{purple}"semantic\_context"}}}{18}
      {"service_description"}{{{\color{purple}"service\_description"}}}{21}
      {"security_profile"}{{{\color{purple}"security\_profile"}}}{18}
      {"service_continuity"}{{{\color{purple}"service\_continuity"}}}{20}
      {"edge_processing"}{{{\color{purple}"edge\_processing"}}}{17},
}

\lstdefinestyle{trislaschema}{
    language=json,
    basicstyle=\ttfamily\scriptsize,
    breaklines=true,
    breakatwhitespace=false,
    columns=fullflexible,
    keepspaces=true,
    showstringspaces=false,
    frame=single,
    framerule=0.5pt,
    rulecolor=\color{gray!30},
    backgroundcolor=\color{LightGray!30},
    xleftmargin=3mm,
    xrightmargin=3mm,
    aboveskip=3pt,
    belowskip=3pt,
    lineskip=-0.5pt,
    tabsize=2,
    captionpos=b
}

\title{
TriSLA: A Preventive and Closed-Loop SLA-Aware Architecture for Multidomain Decision-Making with Explainable Artificial Intelligence in 5G Networks
}
\author{Abel~J.~R.~Lisboa,
        Gustavo~Z.~Bruno\orcidlink{0000-0002-1424-3404},
        and~Cristiano~B.~Both\orcidlink{0000-0002-9776-4888},~\IEEEmembership{Member,~IEEE}%
\thanks{Abel~J.~R.~Lisboa and Cristiano~B.~Both are with the Applied Computing Graduate Program, Universidade do Vale do Rio dos Sinos (UNISINOS), S\~{a}o Leopoldo 93022-750, Brazil (e-mail: abell@edu.unisinos.br; cbboth@unisinos.br).}%
\thanks{Gustavo~Z.~Bruno is with the Instituto Nacional de Telecomunica\c{c}\~{o}es (Inatel), Santa Rita do Sapuca\'{\i} 37540-000, Brazil (e-mail: gustavo.zanatta@posdoc.inatel.br).}%
}

\begin{document}

\maketitle

% =========================================================
% ABSTRACT
% =========================================================

\input{sections/1-Abstract}

% =========================================================
% KEYWORDS
% =========================================================

\acresetall

% =========================================================
% MAIN SECTIONS
% =========================================================

\input{sections/2-Introduction}

\input{sections/3-Background}

\input{sections/4-RelatedWorks}

\input{sections/5-Architecture}

\input{sections/6-Prototype}

\input{sections/7-Methodology}

\input{sections/8-Results}

\input{sections/9-Conclusion}

\input{sections/9-Acknowledgment}

% =========================================================
% REFERENCES
% =========================================================
\bibliographystyle{IEEEtran}
\bibliography{references}

% =========================================================
% BIOGRAPHIES
% =========================================================
\input{sections/10-Biographies}

% =========================================================
% END DOCUMENT
% =========================================================
\end{document}

%% file: acronyms.tex
\acrodef{5GC}{5G Core}
\acrodef{AF}{Application Function}
\acrodef{AMF}{Access and Mobility Management Function}
\acrodef{AI}{Artificial Intelligence}
\acrodef{API}{Application Programming Interface}
\acrodef{CSMF}{Communication Service Management Function}
\acrodef{DRL}{Deep Reinforcement Learning}
\acrodef{E2E}{End-to-End}
\acrodef{eMBB}{Enhanced Mobile Broadband}
\acrodef{LSTM}{Long Short-Term Memory}
\acrodef{MEC}{Multi-access Edge Computing}
\acrodef{mMTC}{Massive Machine-Type Communications}
\acrodef{MLP}{Multi-Layer Perceptron}
\acrodef{NEF}{Network Exposure Function}
\acrodef{NLP}{Natural Language Processing}
\acrodef{NRF}{Network Repository Function}
\acrodef{NSMF}{Network Slice Management Function}
\acrodef{NSSF}{Network Slice Selection Function}
\acrodef{NSSMF}{Network Slice Subnet Management Function}
\acrodef{O-RU}{Open Radio Unit}
\acrodef{PCF}{Policy Control Function}
\acrodef{QoD}{Quality on Demand}
\acrodef{QoE}{Quality of Experience}
\acrodef{QoS}{Quality of Service}
\acrodef{RAN}{Radio Access Network}
\acrodef{SMF}{Session Management Function}
\acrodef{SMO}{Service Management and Orchestration}
\acrodef{TN}{Transport Network}
\acrodef{UPF}{User Plane Function}
\acrodef{UE}{User Equipment}
\acrodef{URLLC}{Ultra-Reliable Low-Latency Communications}
\acrodef{VR}{Virtual Reality}
\acrodef{XAI}{Explainable Artificial Intelligence}
\acrodef{gNB}{next-generation NodeB}
\acrodef{SLA}{Service Level Agreement}
\acrodef{GST}{Generic Slice Template}
\acrodef{NEST}{Network Slice Template}
\acrodef{SDN}{Software-Defined Networking}
\acrodef{ML}{Machine Learning}
\acrodef{NASP}{Network Slice as a Service Platform}
\acrodef{NSaaS}{Network Slice as a Service}
\acrodef{GSMA}{GSM Association}
\acrodef{PRB}{Physical Resource Block}
\acrodef{SHAP}{SHapley Additive exPlanations}
\acrodef{ZSM}{Zero-touch network and Service Management}
\acrodef{ETSI}{European Telecommunications Standards Institute}
\acrodef{3GPP}{3rd Generation Partnership Project}
\acrodef{SST}{Slice/Service Type}
\acrodef{SD}{Slice Differentiator}
\acrodef{NFV}{Network Functions Virtualization}
\acrodef{SBI}{Service-Based Interface}
\acrodef{NGAP}{Next Generation Application Protocol}
\acrodef{GTP-U}{GPRS Tunneling Protocol User Plane}
\acrodef{PDU}{Protocol Data Unit}

%% file: sections/1-Abstract.tex
\begin{abstract}

Network slicing in multidomain 5G environments introduces critical challenges in guaranteeing Service Level Agreements (SLAs) under dynamic resource variability and heterogeneous service requirements.
This article presents TriSLA, a closed-loop, preventive, SLA-aware architecture designed to evaluate feasibility at request time and continuously ensure SLA compliance during operation.
The architecture combines ontology-driven semantic intent interpretation, multidomain machine learning feasibility risk inference, Explainable Artificial Intelligence (XAI) feature attribution, and closed-loop runtime SLA assurance into a unified operational pipeline.
A fully operational prototype was evaluated in a multi-node cloud-native environment integrating Radio Access Network (RAN), Transport Network (TN), and 5G Core (5GC) domains with real-time telemetry collection.
Experimental evaluation demonstrates that TriSLA guarantees a 100\% SLA satisfaction rate for admitted slices, completely eliminating post-deployment violations compared to reactive (51.2\%) and static threshold (80.4\%) admission baselines.
The predictive feasibility assessment achieved a classification accuracy of up to 99.51\% (98.68\% for the default explainable Random Forest classifier), enabling preventive admission decisions before infrastructure commitment.
Furthermore, the cognitive admission pipeline introduces minimal processing overhead, requiring 25.37\,ms for ontology-driven semantic parsing and 231.66\,ms for XAI-assisted feasibility inference.
Concurrently, the closed-loop assurance engine resolves 100\% of runtime telemetry anomalies within a 4.22\,s recovery cycle.
These results demonstrate that TriSLA provides reliable, explainable, transparent, and preventive SLA management through integrated predictive admission and closed-loop runtime assurance for next-generation 5G networks.
\end{abstract}

%% file: sections/2-Introduction.tex
% =========================
\section{Introduction}
\label{sec:introduction}
% =========================

% --- Paragraph 1: High-Level Context (5G Network Slicing & Service Heterogeneity) ---
The evolution of fifth-generation (5G) mobile networks and network slicing has transformed network orchestration~\cite{Kaloxylos2018Slicing, Afolabi2018Network}.
This transformation changes the way resources are allocated, orchestrated, and supervised~\cite{Kaloxylos2018Slicing}.
Network slicing enables multiple logical networks to coexist over a shared physical infrastructure while supporting heterogeneous service categories such as \ac{URLLC}, \ac{eMBB}, and \ac{mMTC}~\cite{Afolabi2018Network, 3GPP2023TS23501}.
Although this coexistence improves flexibility and service customization, it also increases the complexity of guaranteeing \acp{SLA} in dynamic multidomain environments.

% --- Paragraph 2: Problem Statement (Multidomain SLA Dependencies & E2E Consistency) ---
The \ac{SLA} compliance depends on the joint behavior of \ac{RAN}, \ac{TN}, and \ac{5GC} domains.
These domains interact under changing infrastructure conditions~\cite{Afolabi2018Network, Li2021SLA}.
Variations in latency, resource allocation, congestion, and instability may compromise service guarantees.
This instability can occur even when individual components appear locally available or adequately provisioned~\cite{Kaloxylos2018Slicing, 3GPP2023TS28541}.
As network slicing becomes increasingly distributed across multiple domains, maintaining end-to-end \ac{SLA} consistency in heterogeneous networks becomes substantially more challenging.

% --- Paragraph 3: State of the Art & Reactive Limitations (Current SLA Management Workflows) ---
Most existing \ac{SLA} management approaches still operate predominantly through reactive workflows~\cite{Li2021SLA, Coronado2022SLA}.
Monitoring systems identify violations only after service degradation has already occurred.
For example, monitoring latency limits the ability of orchestration platforms to prevent infeasible slice deployments at admission time~\cite{Bega2021Admission, Sun2022SLAAdmission, Sciancalepore2019RLNSB}.
While these approaches improve operational visibility, they do not guarantee that accepted \acp{SLA} remain sustainable in mission-critical scenarios where delayed reactions compromise service continuity~\cite{Popovski2018URLLC, 3GPP2023TS23501}.
Overcoming these limitations requires adopting a preventive closed-loop \ac{SLA} management paradigm.
This paradigm shifts \ac{SLA} assurance from post-deployment violation detection to pre-instantiation predictive admission control integrated with continuous runtime telemetry feedback.
To address these challenges, recent research leverages \ac{AI} techniques to improve orchestration efficiency, resource allocation, and adaptive decision-making~\cite{Shen2020Orchestration, Njah2025AI, Ochonu2024AI}.
Furthermore, intent-driven management paradigms and standardized semantic models, such as \acp{GST}, allow tenants to express high-level service intents without configuring low-level network details~\cite{GSMA2021NG116, Leivadeas2023Intent}.

% --- Paragraph 4: Research Gap Analysis (Four Interdependent Limitations) ---
Despite these advances, achieving end-to-end \ac{SLA} assurance remains an open problem due to interdependent challenges.
First, semantic intent models lack formal ontology-driven validation and deterministic fallback mapping across heterogeneous technical domains, preventing the reliable translation of ambiguous service intents into machine-enforceable operational profiles.
Second, feasibility decisions made at request time fail to align joint runtime infrastructure conditions across the \ac{RAN}, \ac{TN}, and \ac{5GC} domains prior to resource commitment~\cite{Kaloxylos2018Slicing, 3GPP2023TS28541, Polese2023ORAN}.
Third, many \ac{AI}-driven orchestration platforms rely on black-box decision models that conceal multidomain feature dependencies, resulting in opaque decisions that hinder operational trust and root-cause analysis~\cite{Adadi2018XAI, Brik2024XAI}.
Without \ac{XAI} mechanisms to expose underlying feature attributions across the \ac{RAN}, \ac{TN}, and \ac{5GC} domains, network operators cannot audit which infrastructure bottlenecks drove an admission rejection or profile renegotiation.
Fourth, existing platforms assume closed-loop behavior without validating whether orchestration, runtime telemetry, lifecycle continuity, and admission recomputation remain causally connected during operation~\cite{ETSI2022ZSM002}.
Consequently, current solutions lack integrated preventive architectures capable of unified cognitive admission, execution monitoring, and closed-loop validation.

% --- Paragraph 5: Proposed Solution & Contributions (TriSLA Architecture & Key Research Contributions) ---
To address these limitations, we propose \textit{TriSLA}, a preventive multidomain architecture designed to evaluate \ac{SLA} feasibility before slice instantiation while maintaining alignment with runtime conditions.
\textit{TriSLA} unifies four complementary pillars addressing these challenges:
(i)~hybrid \ac{NLP} and ontology-driven semantic processing to resolve intent ambiguity and construct standardized operational profiles using formal slice schemas~\cite{GSMA2021NG116};
(ii)~multidomain predictive inference to evaluate slice feasibility against real-time \ac{RAN}, \ac{TN}, and \ac{5GC} telemetry prior to resource commitment~\cite{Njah2025AI, Ochonu2024AI};
(iii)~\ac{XAI} decision arbitration to generate transparent multidomain feature attributions for trustworthy cognitive admission; and
(iv)~runtime telemetry integration for continuous monitoring and closed-loop remediation upon performance drift.
In this context, the main contributions of this work are summarized as follows:
\begin{itemize}
    \item The design of \textit{TriSLA}, a preventive multidomain \ac{SLA}-aware architecture that unifies intent interpretation, predictive feasibility evaluation, transparent admission, and closed-loop supervision into an integrated 5G slicing architecture.
    \item An ontology-driven semantic interpretation pipeline combining \ac{NLP} extraction with formal domain ontologies to map ambiguous service intents into standardized, machine-enforceable operational profiles.
    \item A multidomain predictive inference engine that correlates real-time telemetry across \ac{RAN}, \ac{TN}, and \ac{5GC} domains to forecast \ac{SLA} satisfaction and filter out infeasible slice requests before resource instantiation.
    \item An \ac{XAI}-assisted decision-making mechanism that computes low-latency feature attributions to provide explainable root-cause rationale for cognitive admission and profile renegotiation.
    \item A runtime telemetry assurance framework enabling continuous monitoring and closed-loop remediation upon performance drift, backed by empirical testbed validation.
\end{itemize}

% --- Paragraph 6: Experimental Validation & Key Results (NASP-based Testbed & Quantitative Metrics) ---
We validate \textit{TriSLA} through an operational prototype deployed on a cloud-native simulated environment extending \ac{NASP}~\cite{Grings2026NASP}.
This cloud-simulated environment orchestrates software emulators across the \ac{RAN} (UERANSIM), \ac{TN} (ONOS and Mininet), and \ac{5GC} (free5GC) domains within a multi-node Kubernetes cluster.
This setup enables high-frequency telemetry collection and reproducible multidomain evaluation under controlled variations in radio capacity, \ac{TN} \ac{SDN} link bottlenecks, and \ac{5GC} workload stress.

Experimental results confirm that multidomain preventive telemetry correlation effectively realizes the \ac{SLA} protection premise prior to resource allocation.
The default Random Forest classifier achieves an \ac{SLA} feasibility classification accuracy of $98.68\%$ and maintains a $100\%$ \ac{SLA} satisfaction rate for admitted slices across heterogeneous service categories.
This preventive filtering eliminates post-deployment violations, whereas reactive and static capacity baselines reach only $51.2\%$ and $80.4\%$ \ac{SLA} satisfaction.
Furthermore, the cognitive admission stage operates with low sub-second overhead, requiring $25.37$\,ms for semantic processing and $231.66$\,ms for \ac{ML} inference and \ac{XAI} feature attribution.
Meanwhile, the closed-loop supervisor successfully resolves cross-domain runtime anomalies within a $4.22$\,s remediation cycle.

% --- Paragraph 7: Paper Organization (Manuscript Roadmap) ---
The remainder of this article is organized as follows.
Section~\ref{sec:background} and Section~\ref{sec:related} present the theoretical background and review related work.
Section~\ref{sec:trisla_architecture} describes the proposed architecture.
Section~\ref{sec:prototype_implementation} and Section~\ref{sec:evaluation_methodology} detail the prototype implementation and the evaluation methodology.
Section~\ref{sec:results} presents the experimental results and discussion.
Finally, Section~\ref{sec:conclusion} concludes the article.

%% file: sections/3-Background.tex
% =========================
\section{Theoretical Background}
\label{sec:background}
% =========================

This section presents the theoretical foundations supporting multidomain 5G network slice assurance.
We examine the integration of 5G network slicing across \ac{RAN}, \ac{TN}, and \ac{5GC} domains with proactive \ac{SLA} assurance.
Additionally, we examine \ac{NSaaS} paradigms and cloud-simulated orchestration platforms.
Finally, we formalize the preventive closed-loop \ac{SLA} management paradigm as the theoretical foundation for proactive, multidomain slice control.

% =========================
\subsection{5G Network Slicing and Multidomain Orchestration}
% =========================

Network slicing enables the partition of a single physical infrastructure into multiple logical networks to support heterogeneous service requirements~\cite{Afolabi2018Network, Kaloxylos2018Slicing, 3GPP2023TS23501}.
This approach allows operators to instantiate distinct profiles, such as \ac{URLLC}, \ac{eMBB}, and \ac{mMTC}, under specific performance constraints.
These profiles accommodate diverse requirements by isolating resources and customizing traffic management policies for each tenant.
However, partitioning physical resources dynamically across shared physical nodes introduces significant orchestration challenges.

To address these challenges, modern 5G architectures adopt disaggregated, virtualized, and programmable \acp{RAN}~\cite{Kaloxylos2018Slicing, Polese2023ORAN}.
By disaggregating base stations into modular software components, 5G supports flexible deployment strategies and fine-grained resource management.
In these architectures, higher-tier orchestration and intelligent decision-making operate within the \ac{SMO} framework, managing long-term policy refinement, cross-domain optimization, and \ac{SLA} lifecycle automation~\cite{Polese2023ORAN}.
Despite these operational benefits, disaggregation increases the complexity of managing end-to-end services across heterogeneous infrastructures.
Ensuring consistent compliance requires coordinated control across the \ac{RAN}, \ac{TN}, and \ac{5GC} domains under dynamic network conditions~\cite{Kaloxylos2018Slicing, 3GPP2023TS28541, Farrel2024RFC9543}.

% =========================
\subsection{Service Level Agreements in Network Slicing}
% =========================

\acp{SLA} define the expected performance, availability, and quality metrics of network services.
In network slicing, these agreements are represented by specific metrics, including latency, jitter, throughput, packet loss, and resource utilization~\cite{Li2021SLA, Coronado2022SLA, GSMA2021NG116}.
Tenants rely on these indicators to guarantee that their application requirements are met by the slice provider.
Therefore, maintaining these indicators within negotiated thresholds is critical for service quality and provider credibility.

Traditional \ac{SLA} management frameworks rely on reactive monitoring and post-deployment enforcement mechanisms~\cite{Li2021SLA}.
Under this approach, operators detect violations and initiate corrective actions only after service degradation has occurred.
This reactive approach is insufficient for mission-critical slices where resource availability must be guaranteed at the moment of instantiation.
Furthermore, dynamic conditions across the \ac{RAN}, \ac{TN}, and \ac{5GC} domains make static, reactive admission decisions highly unreliable~\cite{Kaloxylos2018Slicing, ETSI2022ZSM002}.

% =========================
\subsection{Explainable Artificial Intelligence for Network Decision-Making}
% =========================

Operators integrate \ac{AI} into orchestration platforms to manage the complexity of multidomain slicing~\cite{Shen2020Orchestration}.
\ac{ML} models process runtime telemetry to predict resource demands and dynamically optimize allocation.
In \ac{SLA}-aware environments, these models estimate service sustainability by correlating resource metrics with requested thresholds~\cite{Njah2025AI}.
By evaluating these patterns, the system can estimate feasibility before committing physical infrastructure resources~\cite{Ochonu2024AI}.

Despite their predictive power, many \ac{AI}-based orchestration systems operate as black-box models.
These models generate decisions without exposing the underlying logic or the feature contributions that led to the outcome.
In mission-critical networks, this opacity reduces operational trust and hampers troubleshooting during service degradation.
Consequently, operators require interpretable methods to validate automated decisions before applying them to the infrastructure~\cite{Ochonu2024AI, Adadi2018XAI}.

\ac{XAI} addresses the limitations of closed black-box models by transforming opaque predictions into open, interpretable decision outcomes~\cite{Adadi2018XAI}.
Methods such as feature-importance analysis and contribution tracking identify which operational variables most affect predictions.
In \ac{SLA} admission workflows, \ac{XAI} translates raw network metrics into explicit contributors to the feasibility decision~\cite{Brik2024XAI}.
This translation allows operators to verify how \ac{RAN}, \ac{TN}, and \ac{5GC} conditions affect service sustainability.
Furthermore, explainability improves operational trust, compliance auditing, and troubleshooting in automated orchestration pipelines~\cite{Brik2024XAI}.

% =========================
\subsection{Network Slice as a Service}
% =========================

\ac{NSaaS} delivers end-to-end network slicing capabilities on demand as a service model, allowing vertical tenants to instantiate and manage customized logical networks with negotiated \ac{SLA} constraints~\cite{3GPP2023TS28531}.
To realize \ac{NSaaS} in practice, multidomain orchestration platforms rely on high-fidelity simulation environments to evaluate lifecycle operations under realistic traffic and resource conditions.
Platforms such as \ac{NASP}~\cite{Grings2026NASP} operationalize \ac{NSaaS} by containerizing network emulators within Kubernetes environments, integrating disaggregated \ac{RAN}, \ac{TN} \ac{SDN} controllers, and \ac{5GC} network functions into a unified operational workflow.
This platform architecture enables programmatic slice provisioning, fine-grained telemetry extraction, and controlled stress injection across heterogeneous network domains.

% =========================
\subsection{Runtime-Aware SLA Management Paradigms}
% =========================

\begin{figure}[!t]
\centering
\includegraphics[width=\linewidth]{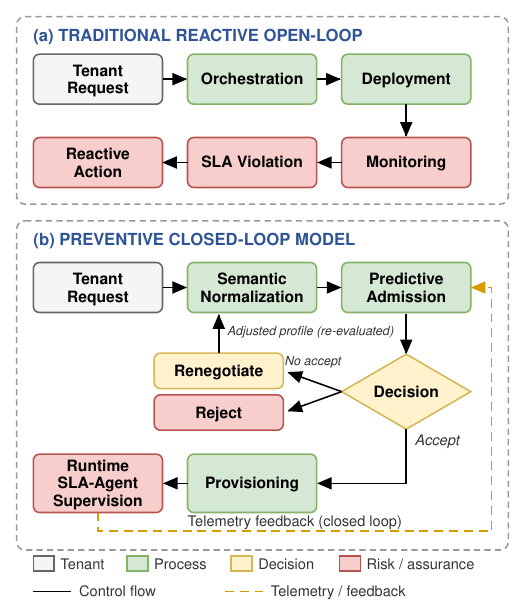}
\caption{Comparison between traditional reactive open-loop SLA management and the preventive closed-loop SLA management model.}
\label{fig:sla_comparison}
\end{figure}

As shown in Fig.~\ref{fig:sla_comparison}(a), traditional open-loop \ac{SLA} management processes tenant requests through orchestration and deployment before monitoring detects \ac{SLA} violations to trigger reactive mitigation~\cite{Li2021SLA, Coronado2022SLA}.
This reactive approach disconnects initial admission decisions from runtime infrastructure behavior, responding only after service degradation has occurred.
To overcome these limitations, the preventive closed-loop \ac{SLA} management model introduces a proactive theoretical framework, shown in Fig.~\ref{fig:sla_comparison}(b).
In this preventive closed-loop model, incoming tenant requests first undergo semantic normalization followed by predictive admission to evaluate feasibility prior to instantiation~\cite{Leivadeas2023Intent}.
The evaluation produces an explicit decision outcome: accepted requests proceed to provisioning and runtime supervision, whereas unaccepted requests branch into renegotiation or rejection.
When renegotiation is selected, an adjusted slice profile returns to semantic normalization for re-evaluation, while runtime supervision continuously feeds operational telemetry back into predictive admission to maintain a closed loop.

Implementing this closed-loop cycle requires coordinating telemetry observability, lifecycle persistence, and operational supervision.
However, maintaining causal relationships between preventive admission and runtime re-evaluation remains a significant challenge.
Most existing platforms fail to integrate telemetry reasoning, orchestration, and runtime validation into a single framework.
Consequently, the theoretical boundaries between admission control and active slice supervision are rarely characterized in practice.

% =========================
\subsection{Synthesis and Research Gap}
% =========================

Although advancements have occurred across individual domains, existing solutions remain fragmented and decoupled.
Current \ac{SLA} management relies heavily on reactive monitoring, whereas \ac{AI}-based orchestration often lacks explainability and semantic validation~\cite{Li2021SLA, Ochonu2024AI}.
Furthermore, while cloud-simulated \ac{NSaaS} platforms like \ac{NASP}~\cite{Grings2026NASP} enable multidomain execution, cross-domain coordination across \ac{RAN}, \ac{TN}, and \ac{5GC} remains poorly integrated with pre-deployment feasibility reasoning.
Specifically, the fundamental research gap is the lack of a unified preventive closed-loop model that evaluates the feasibility of joint multidomain \ac{SLA} prior to slice instantiation, while integrating transparent decision explainability and continuous runtime telemetry supervision.
Consequently, current frameworks struggle to unify semantic interpretation, predictive admission, and runtime validation within a single closed-loop lifecycle.
The operational boundaries separating preventive admission, orchestration, and runtime re-evaluation remain insufficiently characterized.
This theoretical gap motivates the design of \textit{TriSLA}, a preventive, closed-loop, \ac{SLA}-aware architecture presented in Section~\ref{sec:trisla_architecture}.
The proposed architecture operationalizes these theoretical concepts by evaluating feasibility before slice instantiation, using integrated semantic processing and multidomain telemetry observability.

%% file: sections/4-RelatedWorks.tex
% =========================
\section{Related Work}
\label{sec:related}
% =========================

\input{Tables/related_work_comparison}

Recent research extensively explores \ac{SLA} management, multidomain orchestration, and intelligent resource allocation in 5G and beyond network slicing.
To provide a structured review of the state of the art, existing literature is categorized into architectural standards and survey taxonomies, followed by technical proposals for slice admission and \ac{SLA} management.
Finally, we synthesize these developments to highlight the operational gaps addressed by the \textit{TriSLA} architecture.

\subsection{Surveys, Standardization, and Specifications}

Standardization bodies and broad literature reviews establish foundational taxonomies and service templates for network slicing.
The \ac{GSMA} NG.116 specification defines the \ac{GST} to standardize \ac{SLA} attribute requirements across telecommunication domains~\cite{GSMA2021NG116}.
Based on these specifications, comprehensive surveys summarize architectural frameworks and management paradigms for the slicing of 5G networks~\cite{Kaloxylos2018Slicing}.
More recently, tutorials and surveys on \ac{XAI} and disaggregated Open \ac{RAN} architectures summarize interpretability techniques and controller specifications applicable to 6G communication management~\cite{Polese2023ORAN, Brik2024XAI, Sun2025XAI, Adadi2018XAI}.
Although these contributions establish terminology and conceptual taxonomies, they focus on static specifications or post hoc analyses.
Consequently, existing surveys and standards do not provide real-time decision algorithms or dynamic workflows that can evaluate \ac{SLA} feasibility before resource commitment.

\subsection{Proposals for SLA Orchestration and Admission Control}

Technical proposals address specific execution stages within the network slice lifecycle, focusing on \ac{SLA}-aware provisioning, intelligent admission control, or multidomain execution.
Early work on \ac{SLA}-aware orchestration emphasizes service specification, dynamic monitoring, and post-deployment validation~\cite{Coronado2022SLA, Li2021SLA}.
Although these orchestration frameworks enable automated management, they rely on reactive feedback loops that detect \ac{SLA} violations only after the resources have already been instantiated.
To mitigate post-deployment failures, optimization-based and \ac{AI}-driven slice admission control models have been introduced~\cite{Bega2021Admission, Ochonu2024AI}.
These admission control models leverage \ac{DRL}, predictive modeling, and federated optimization to regulate the acceptance of slice requests~\cite{Shen2020Orchestration, Xu2020AIAdmission, Sciancalepore2019RLNSB}.
However, most intelligent admission models operate as opaque decision systems, limiting operational trust in mission-critical environments where rejected or miscalculated slice requests disrupt service continuity.

Semantic service modeling aims to map business-level intent into structured network configurations through high-level ontologies and intent-based management frameworks~\cite{Njah2025AI, Leivadeas2023Intent}.
Nevertheless, semantic profiles are applied as static provisioning models rather than active participants in dynamic admission workflows.
From an infrastructure perspective, multidomain \ac{NSaaS} platforms such as \ac{NASP} demonstrate progress toward unified orchestration and continuous validation across \ac{RAN}, \ac{TN}, and \ac{5GC} domains~\cite{Grings2026NASP}.
Despite these advances, existing multidomain platforms focus predominantly on post-activation adaptation rather than pre-deployment assurance.
Preventive \ac{SLA} validation before infrastructure commitment remains unaddressed, particularly when semantic intent translation, explainable reasoning, and real-time multidomain telemetry must operate synchronously.

\subsection{Comparative Analysis and Research Gap}

Table~\ref{tab:related_work_comparison} categorizes the representative state-of-the-art literature into four distinct technical domains (surveys and standards, \ac{AI}-driven admission control, \ac{SLA} and intent orchestration, and multidomain platforms) and compares them against \textit{TriSLA}.
The evaluation criteria assess seven key capabilities: preventive admission control (Prev.), multidomain coverage across \ac{RAN}, \ac{TN}, and \ac{5GC} (R/T/C), \ac{AI} decision logic (AI), explainability (XAI), semantic modeling (Ont.), closed-loop execution (CL), and testbed validation (Real, where partial support denotes simulated or emulated prototypes vs. commercial hardware).
Additionally, the table outlines primary operational limitations to highlight integration gaps in the existing literature.

As shown in Table~\ref{tab:related_work_comparison}, existing works cover only specific subsets of these requirements.
Survey articles and specifications establish service templates and management taxonomies without executing runtime decisions~\cite{Kaloxylos2018Slicing, GSMA2021NG116, Polese2023ORAN, Brik2024XAI, Sun2025XAI}.
Algorithmic proposals introduce predictive admission or intent modeling, but rely on opaque decision logic or isolated simulations that omit complex multidomain interactions and container orchestration overheads~\cite{Bega2021Admission, Xu2020AIAdmission, Sciancalepore2019RLNSB, Ochonu2024AI, Njah2025AI, Leivadeas2023Intent}.
Furthermore, multidomain execution platforms lack pre-deployment preventive admission control guided by explainable feedback~\cite{Grings2026NASP}.

These research gaps motivate \textit{TriSLA}, an integrated architecture designed for the preventive management of \ac{SLA}-aware slices.
\textit{TriSLA} coordinates ontology-assisted semantic interpretation with explainable \ac{ML} inference before resource allocation.
Furthermore, the architecture integrates real-time telemetry correlation across disaggregated domains with closed-loop execution.
By validating feasibility before committing to infrastructure on a cloud-native prototype testbed, \textit{TriSLA} bridges the gap between pre-deployment admission control and continuous runtime assurance.

%% file: Tables/related_work_comparison.tex
\begin{table*}[!t]
\centering
\caption{Comparison between related work and TriSLA.}
\label{tab:related_work_comparison}

\renewcommand{\arraystretch}{1.25}
\resizebox{\textwidth}{!}{%
\begin{tabular}{l lll ccccccc l}
\toprule
\rowcolor{LightGray!100}
& \multicolumn{3}{c}{\textbf{Article Details}} & \multicolumn{7}{c}{\textbf{Key Capabilities}} & \textbf{Limitation} \\ \cmidrule(lr){2-4} \cmidrule(lr){5-11} \cmidrule(lr){12-12}
\rowcolor{LightGray!100}
\textbf{Category} & \textbf{Work} & \textbf{Year} & \textbf{Focus} & \textbf{Prev.} & \textbf{R/T/C} & \textbf{AI} & \textbf{XAI} & \textbf{Ont.} & \textbf{CL} & \textbf{Real} & \\ \midrule

\rowcolor{white}
& Kaloxylos~\cite{Kaloxylos2018Slicing} & 2018 & Slicing survey & \EmptyCircle & \EmptyCircle & \EmptyCircle & \EmptyCircle & \EmptyCircle & \EmptyCircle & \EmptyCircle & No validation \\
\rowcolor{LightGray!40}
& GSMA~\cite{GSMA2021NG116} & 2021 & SLA templates & \HalfCircle & \HalfCircle & \EmptyCircle & \EmptyCircle & \FullCircle & \EmptyCircle & \EmptyCircle & No runtime decision \\
\rowcolor{white}
& Polese~\cite{Polese2023ORAN} & 2023 & O-RAN architecture & \EmptyCircle & \HalfCircle & \HalfCircle & \EmptyCircle & \EmptyCircle & \HalfCircle & \EmptyCircle & Survey only \\
\rowcolor{LightGray!40}
& Brik~\cite{Brik2024XAI} & 2024 & XAI O-RAN survey & \EmptyCircle & \HalfCircle & \HalfCircle & \FullCircle & \EmptyCircle & \EmptyCircle & \EmptyCircle & No admission \\
\rowcolor{white}
\multirow{-5}{*}{\begin{tabular}[l]{@{}l@{}}\textbf{Surveys \&}\\\textbf{Standards}\end{tabular}}
& Sun~\cite{Sun2025XAI} & 2025 & XAI slicing survey & \EmptyCircle & \EmptyCircle & \HalfCircle & \FullCircle & \EmptyCircle & \EmptyCircle & \EmptyCircle & No validation \\ \midrule

\rowcolor{LightGray!40}
& Sciancalepore~\cite{Sciancalepore2019RLNSB} & 2019 & DRL slice broker & \FullCircle & \HalfCircle & \FullCircle & \EmptyCircle & \EmptyCircle & \HalfCircle & \EmptyCircle & No XAI \\
\rowcolor{white}
& Bega~\cite{Bega2021Admission} & 2020 & AI slice mgmt & \HalfCircle & \HalfCircle & \FullCircle & \EmptyCircle & \EmptyCircle & \EmptyCircle & \EmptyCircle & Simulation focus \\
\rowcolor{LightGray!40}
& Shen~\cite{Shen2020Orchestration} & 2020 & AI-assisted slicing & \HalfCircle & \HalfCircle & \FullCircle & \EmptyCircle & \EmptyCircle & \HalfCircle & \EmptyCircle & No real validation \\
\rowcolor{white}
& Abdellatif~\cite{Xu2020AIAdmission} & 2023 & AI admission & \FullCircle & \HalfCircle & \FullCircle & \EmptyCircle & \EmptyCircle & \HalfCircle & \EmptyCircle & Opaque decision \\
\rowcolor{LightGray!40}
\multirow{-5}{*}{\begin{tabular}[l]{@{}l@{}}\textbf{AI Admission}\\\textbf{Control}\end{tabular}}
& Ochonu~\cite{Ochonu2024AI} & 2024 & Slice-aware AC & \FullCircle & \HalfCircle & \HalfCircle & \EmptyCircle & \EmptyCircle & \HalfCircle & \EmptyCircle & Factory domain only \\ \midrule

\rowcolor{white}
& Li~\cite{Li2021SLA} & 2021 & SLA provisioning & \HalfCircle & \HalfCircle & \EmptyCircle & \EmptyCircle & \EmptyCircle & \HalfCircle & \HalfCircle & No admission \\
\rowcolor{LightGray!40}
& Coronado~\cite{Coronado2022SLA} & 2022 & ZSM automation & \EmptyCircle & \HalfCircle & \HalfCircle & \EmptyCircle & \EmptyCircle & \HalfCircle & \EmptyCircle & No SLA focus \\
\rowcolor{white}
& Leivadeas~\cite{Leivadeas2023Intent} & 2023 & IBN survey & \EmptyCircle & \HalfCircle & \FullCircle & \EmptyCircle & \HalfCircle & \HalfCircle & \EmptyCircle & No SLA focus \\
\rowcolor{LightGray!40}
\multirow{-4}{*}{\begin{tabular}[l]{@{}l@{}}\textbf{SLA \& Intent}\\\textbf{Orchestration}\end{tabular}}
& Njah~\cite{Njah2025AI} & 2025 & Intent-based arch. & \EmptyCircle & \HalfCircle & \FullCircle & \EmptyCircle & \HalfCircle & \HalfCircle & \EmptyCircle & No SLA focus \\ \midrule

\rowcolor{white}
\begin{tabular}[l]{@{}l@{}}\textbf{Multidomain}\\\textbf{Platforms}\end{tabular}
& Grings~\cite{Grings2026NASP} & 2026 & NASP platform & \EmptyCircle & \FullCircle & \EmptyCircle & \EmptyCircle & \EmptyCircle & \HalfCircle & \HalfCircle & No SLA decision \\ \midrule

\rowcolor{LightGray!100}
\begin{tabular}[l]{@{}l@{}}\textbf{Proposed}\\\textbf{Framework}\end{tabular}
& \textbf{TriSLA (this work)} & \textbf{2026} & \textbf{Closed-loop SLA decision} & \FullCircle & \FullCircle & \FullCircle & \FullCircle & \FullCircle & \FullCircle & \HalfCircle & \textbf{Integrated architecture} \\ \bottomrule

\end{tabular}
}

\vspace{3pt}
\centering\footnotesize
\FullCircle\,supported\enspace
\HalfCircle\,partially supported\enspace
\EmptyCircle\,not supported

\end{table*}

%% file: sections/5-Architecture.tex
% =========================
\section{TriSLA Architecture}
\label{sec:trisla_architecture}
% =========================

\begin{figure}[!t]
\centering
\includegraphics[width=\linewidth]{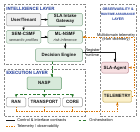}
\caption{\textit{TriSLA} reference architecture integrating semantic admission, multidomain orchestration, and runtime SLA assurance.}
\label{fig:trisla_architecture}
\end{figure}

\textit{TriSLA} is a closed-loop preventive \ac{SLA} assurance architecture designed to evaluate slice feasibility prior to infrastructure resource commitment.
The architecture connects semantic intent requests to operational slice configurations across the \ac{RAN}, \ac{TN}, and \ac{5GC} domains by unifying semantic interpretation, predictive admission, and closed-loop supervision.
By validating service requirements against real-time infrastructure conditions, \textit{TriSLA} prevents \ac{SLA} violations before triggering domain orchestration.
As illustrated in Fig.~\ref{fig:trisla_architecture}, the reference architecture is organized into three core functional tiers: the Intelligence Layer, the Execution Layer, and the Observability and Runtime Assurance Layer, interconnected through standardized interface contracts.
Its microservice design extends \ac{3GPP}, \ac{ETSI} \ac{ZSM}, \ac{GSMA} \ac{NEST}, and \ac{NASP} specifications~\cite{3GPP2023TS23501, 3GPP2023TS28541, ETSI2022ZSM002, GSMA2021NG116, Grings2026NASP} with preventive admission control and \ac{XAI}.
Logically aligned with standard \ac{SMO} frameworks~\cite{Polese2023ORAN}, \textit{TriSLA} operates at the management plane to coordinate admission decisions and continuous runtime assurance across disaggregated network domains.

Each functional tier hosts dedicated components tailored to specific phases of the slice lifecycle, as shown in Fig.~\ref{fig:trisla_architecture}.
The Intelligence Layer hosts the \ac{SLA} Intake Gateway, SEM-\ac{CSMF}, ML-\ac{NSMF}, and the Decision Engine, executing semantic intent normalization, predictive admission inference, and multidomain feasibility arbitration.
The Execution Layer incorporates the \ac{NASP} Adapter and southbound domain controllers across \ac{RAN}, \ac{TN}, and \ac{5GC}, translating validated admission decisions into coordinated provisioning actions across domain infrastructure.
The Observability and Runtime Assurance Layer comprises the multidomain telemetry aggregation component and the \ac{SLA}-Agent, which ingest cross-domain metrics and enforce closed-loop verification to detect performance drift.
The following subsections detail the internal design of the Intelligence Layer (Section~\ref{subsec:intelligence_layer}), the Execution Layer (Section~\ref{subsec:execution_layer}), the Observability and Runtime Assurance Layer (Section~\ref{subsec:observability_runtime_assurance}), and their integrated interaction workflow (Section~\ref{subsec:e2e_workflow}).

% =========================================================
\subsection{Intelligence Layer}
\label{subsec:intelligence_layer}
% =========================================================

This layer evaluates whether the conditions of the multidomain infrastructure can sustain the requested \ac{SLA} before the resources are committed.
Tenants submit service requests via the \ac{SLA} Intake Gateway to the SEM-\ac{CSMF}, which normalizes them into structured semantic profiles categorized by service type, such as \ac{URLLC}, \ac{eMBB}, and \ac{mMTC}.
The ML-\ac{NSMF} is designed as a model-agnostic microservice that exposes a standardized inference interface, allowing the deployment of several \ac{ML} classifiers.
This decoupled inference \ac{API} allows network operators to substitute \ac{ML} back-ends to meet specific computational constraints.
This component correlates normalized service profiles with incoming multidomain telemetry streams to estimate admission feasibility, predictive risk, confidence levels, and explainability metadata.

The Decision Engine evaluates these indicators alongside policy constraints to determine if the service can be sustained in the \ac{RAN}, \ac{TN}, and \ac{5GC} domains.
Rather than performing isolated checks, it assesses the combined multidomain state to converge on an immediate operational outcome: \texttt{ACCEPT}, \texttt{RENEGOTIATE}, or \texttt{REJECT}.
To avoid latency bottlenecks on the critical admission path, the system employs an asynchronous processing pattern.
Immediate admission decisions are returned to the Tenant and forwarded to the execution layer for low-latency provisioning.
Moreover, detailed \ac{SHAP} feature attributions~\cite{Lundberg2020TreeSHAP} execute asynchronously in background tasks, preserving complete explainability metadata without delaying slice instantiation.

% =========================================================
\subsection{Execution Layer}
\label{subsec:execution_layer}
% =========================================================

This layer translates accepted admission decisions into coordinated provisioning workflows across the underlying network infrastructure.
At the top of this tier, the \ac{NASP} Adapter ingests \texttt{ACCEPT} outcomes from the Decision Engine and decomposes them into domain-specific provisioning tasks while respecting inter-domain dependencies.
These provisioning directives are dispatched downward across dedicated southbound interface contracts to the \ac{RAN}, \ac{TN}, and \ac{5GC} controllers, in alignment with standard slicing frameworks~\cite{Farrel2024RFC9543}.
At the domain tier, the respective domain controllers enforce resource reservation and configure local slices across the \ac{RAN}, \ac{TN}, and \ac{5GC} domains.
Upon completing local configuration, domain controllers exchange activation states and acknowledgments with the adapter to verify end-to-end service readiness.
Finally, the adapter publishes deployment outcomes and orchestration metadata to the monitoring subsystem, transitioning the provisioned slice from the admission pipeline to operational runtime management.

%=========================================================
\subsection{Observability and Runtime Assurance Layer}
\label{subsec:observability_runtime_assurance}
% =========================================================

This layer secures operational transparency and continuous assurance through multidomain telemetry ingestion and lifecycle supervision.
As depicted in Fig.~\ref{fig:trisla_architecture}, this layer integrates a dedicated telemetry aggregation component and the \ac{SLA}-Agent.
The telemetry component continuously collects operational metrics from the underlying \ac{RAN}, \ac{TN}, and \ac{5GC} domains, streaming live data to both the ML-\ac{NSMF} for admission risk assessment and the \ac{SLA}-Agent for ongoing tracking.
Operating asynchronously alongside real-time request evaluation, this layer records admission results, provisioning metadata, and runtime performance indicators linked to each service instance.
This decoupled design isolates the telemetry recording from the critical admission decision path while maintaining a complete lifecycle audit record.
Upon slice instantiation, the Decision Engine registers runtime supervision parameters with the \ac{SLA}-Agent, initiating closed-loop verification.
By correlating live telemetry with the committed \ac{SLA} profile, the \ac{SLA}-Agent detects performance drift, compliance deviations, and resource bottlenecks and publishes assurance events back to the monitoring system.

\begin{figure*}[!t]
\centering
\includegraphics[width=\textwidth]{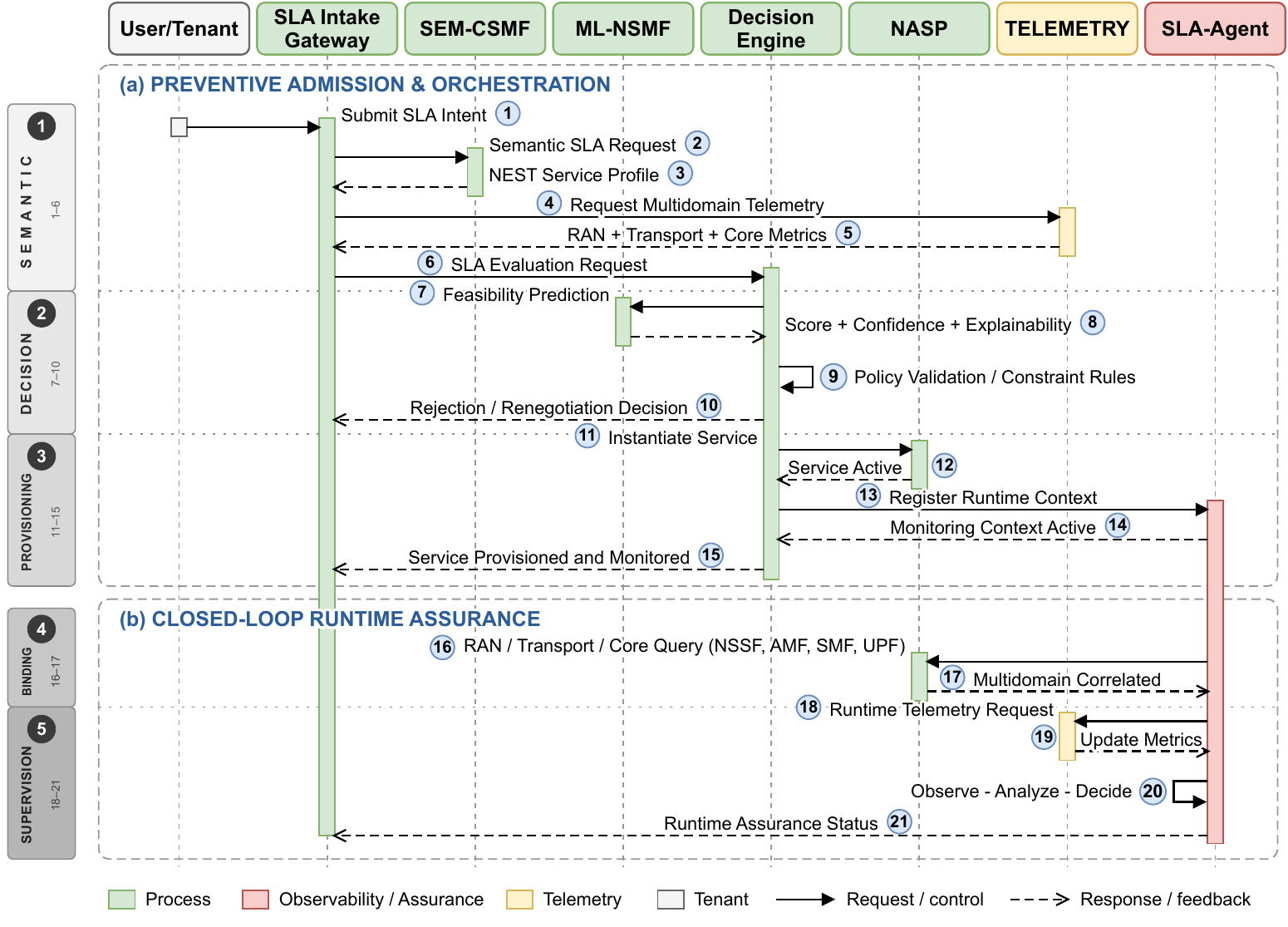}
\caption{\textit{TriSLA} end-to-end SLA admission and orchestration workflow.}
\label{fig:trisla_sequence}
\end{figure*}

% =========================================================
\subsection{End-to-End Runtime Workflow}
\label{subsec:e2e_workflow}
% =========================================================

\textit{TriSLA} coordinates slice admission and lifecycle governance across two primary macro-operational stages, comprising five sequential, closed-loop phases spanning 21 interactions, as illustrated in Fig.~\ref{fig:trisla_sequence}.
Under the first stage, \textit{Preventive Admission and Orchestration}, the architecture evaluates tenant service feasibility and enacts multidomain resource provisioning prior to traffic activation, structured into five sequential benchmarking macro steps (M01--M05).
\begin{itemize}
    \item Phase~1: Semantic Intake and Profiling (Interactions 1--6, Macro Step M01) begins when a tenant submits service intent (1) via the \ac{SLA} Intake Gateway to the SEM-\ac{CSMF} (2).
The SEM-\ac{CSMF} normalizes the request into a canonical \ac{NEST} profile (3), prompting the gateway to retrieve an infrastructure telemetry snapshot (4--5) and dispatch the evaluation request to the Decision Engine (6).

    \item Phase~2: Predictive Decision and Feasibility Arbitration (Interactions 7--10, Macro Step M02) queries the ML-\ac{NSMF} (7) to estimate predictive violation risks, confidence levels, and explainability metadata (8).
The Decision Engine validates local policy rules (9) to produce an immediate outcome of \texttt{ACCEPT}, \texttt{RENEGOTIATE}, or \texttt{REJECT} returned to the gateway (10), while background tasks compute \ac{XAI} attributions.
Following an affirmative admission outcome, the workflow transitions to multidomain resource deployment and observability binding.

    \item Phase~3: Multidomain Provisioning (Interactions 11--15, Macro Step M03) engages the \ac{NASP} Adapter (11), which decomposes the slice intent into atomic configurations across the \ac{RAN}, \ac{TN}, and \ac{5GC} domain controllers.
Upon receiving service activation confirmation (12), the Decision Engine registers runtime supervision parameters with the \ac{SLA}-Agent (13--14) and confirms deployment to the gateway (15).

    \item Phase~4: Observability Context Binding (Interactions 16--17, Macro Step M04) queries network functions via the \ac{NASP} Adapter (16) to bind active slice identifiers across the \ac{RAN}, \ac{TN}, and \ac{5GC} domains, establishing correlated telemetry tracking (17).
The gateway finalizes and returns the deployment confirmation response to the tenant (Macro Step M05).
Under the second stage, \textit{Continuous Runtime Assurance}, the architecture maintains active lifecycle supervision to ensure continuous compliance against agreed service targets.

    \item Phase~5: Continuous SLA Runtime Supervision (Interactions 18--21) continuously requests and updates operational telemetry metrics (18--19) at the \ac{SLA}-Agent.
The \ac{SLA}-Agent executes an Observe-Analyze-Decide control cycle (20) to detect performance drift, resource contention, and emerging compliance anomalies.
Runtime assurance status and anomaly alerts (21) are published to the gateway and management planes to drive automated closed-loop remediation.
This closed-loop lifecycle design guarantees that operational execution remains strictly aligned with the preventive admission decisions committed during initial slice onboarding.
\end{itemize}

The \textit{TriSLA} architecture establishes a unified framework that bridges high-level semantic intents with granular multidomain infrastructure control.
By coupling preventive \ac{ML}-based admission with continuous runtime observability, the architecture prevents \ac{SLA} degradation before commitment while dynamically adapting to operational drift.
%To evaluate the practical feasibility, latency overhead, and assurance efficacy of this architectural design under realistic conditions, Section~\ref{sec:prototype_implementation} details the concrete cloud-native prototype.

%% file: sections/6-Prototype.tex
% =========================
\section{Prototype Implementation}
\label{sec:prototype_implementation}
% =========================

\begin{figure*}[!t]
\centering
\includegraphics[width=\textwidth]{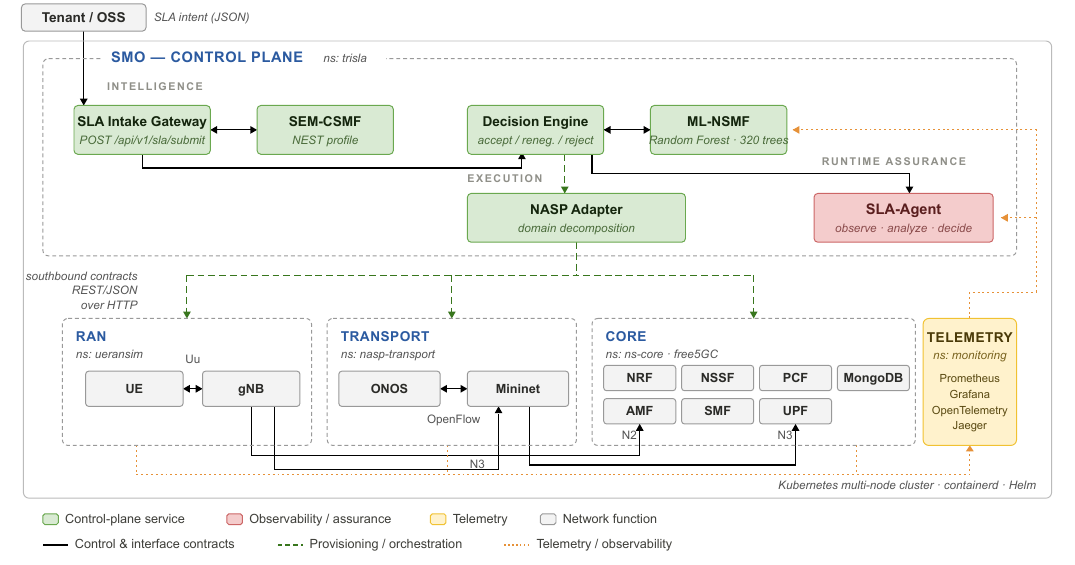}
\caption{\textit{TriSLA} prototype integrating admission control, multidomain orchestration, observability, and runtime assurance services.}
\label{fig:trisla_prototype_topology}
\end{figure*}

This section details the implementation of the \textit{TriSLA} prototype in a multidomain 5G environment.
The control plane implements the \ac{SMO} tier by integrating the \ac{SLA} Intake Gateway, SEM-\ac{CSMF}, ML-\ac{NSMF}, Decision Engine, \ac{NASP} Adapter, and \ac{SLA}-Agent as containerized microservices.
These services execute within Kubernetes pods to support isolated lifecycle management and inter-service communication.
Fig.~\ref{fig:trisla_prototype_topology} shows the prototype components and their communication interfaces across the control and infrastructure planes.

\subsection{Microservice Architecture and Service Orchestration}
\label{subsec:prototype_microservices}

The \textit{TriSLA} control plane deploys as a set of containerized microservices within the dedicated \texttt{trisla} Kubernetes namespace.
Each functional component runs inside isolated pods managed through declarative Kubernetes deployments, services, and namespace policies.
Configuration parameters, resource quotas, and runtime environment variables are injected dynamically into container runtimes via Kubernetes ConfigMaps.
All deployment manifests, Helm charts, configurations, and source code are maintained in the artifact repository at \url{https://github.com/abelisboa/TriSLA}.

Tenant service requests enter through the \ac{SLA} Intake Gateway, which validates input schemas and forwards JSON payloads to the SEM-\ac{CSMF}.
Listing~\ref{lst:semantic_request} shows an example request specifying quantitative performance thresholds alongside semantic descriptors such as service continuity and edge execution.
The SEM-\ac{CSMF} parses these heterogeneous parameters into canonical \ac{NEST} profiles, standardizing admission inputs across slice types.
These normalized profiles decouple tenant-level intent representations from low-level infrastructure requirements during feasibility evaluation.

\begin{lstlisting}[style=trislaschema,
float,
caption={Example semantic SLA request payload processed by \textit{TriSLA}.},
label={lst:semantic_request}]
{
"tenant_id": "tenant-01",
"slice_service_type": "eMBB",
"service_requirements": {
"latency_ms": 10.0,
"throughput_mbps": 100.0,
"availability": 0.999,
"device_density": 1000
  },
"semantic_context": {
"service_description": "VR Streaming",
"security_profile": "high",
"service_continuity": true,
"edge_processing": true
  }
}
\end{lstlisting}

Inter-service communication across the control plane relies on synchronous REST \acp{API} that exchange structured JSON payloads over HTTP.
Internal service discovery and traffic routing are managed through Kubernetes ClusterIP abstractions to ensure reliable endpoint resolution within the cluster.
Declarative Helm charts coordinate release packaging, parameter injection, and compute resource quotas across all deployed control pods.
This modular orchestration allows operators to update or scale individual microservices independently without disrupting active admission or runtime assurance pipelines.

\subsection{Layer-Specific Implementation Details}
\label{subsec:prototype_layers}

The software prototype maps the architectural tiers of \textit{TriSLA} into concrete execution runtimes and communication interfaces.
Each functional layer exposes standardized REST interfaces to separate admission evaluation, resource orchestration, and telemetry supervision.
This separation of concerns ensures that compute-intensive inference and telemetry streaming do not degrade control plane responsiveness.
The following paragraphs describe the specific software stacks and operational mechanisms implemented across each architectural layer.

\textbf{Intelligence Layer.}
This layer integrates the SEM-\ac{CSMF}, ML-\ac{NSMF}, and Decision Engine into the admission pipeline.
Tenants submit JSON requests to the \ac{SLA} Intake Gateway, implemented using FastAPI and Uvicorn, via endpoint \texttt{/api/v1/sla/submit}.
The ML-\ac{NSMF} ingests telemetry snapshots across the \textit{OBS-I1} interface to evaluate admission feasibility using a Random Forest classifier~\cite{Breiman2001RandomForests} trained with scikit-learn.
The model occupies 0.36\,MB with 320 decision trees across 19 features, supporting sub-millisecond inference and \ac{SHAP} attributions.
REST \acp{API} transfer profile attributes, risk estimates, and admission outcomes between the intelligence services.

\textbf{Execution Layer.}
The \ac{NASP} Adapter translates validated admission decisions into structured provisioning workflows for domain execution controllers.
It decomposes end-to-end slice intents into domain-specific lifecycle directives targeting the \ac{RAN}, \ac{TN}, and \ac{5GC} subsystems.
Southbound provisioning requests are dispatched asynchronously across the \textit{RAN-I1}, \textit{TN-I1}, and \textit{CN-I1} interfaces to trigger domain configurations.
This non-blocking dispatch mechanism prevents domain configuration latency from obstructing concurrent admission evaluations in the Decision Engine.

\textbf{Observability and Runtime Assurance Layer.}
This layer implements closed-loop assurance through a distributed telemetry ingestion pipeline and dedicated \ac{SLA}-Agent supervision microservices.
The \textit{SLAA-I1} interface connects the Decision Engine with the \ac{SLA}-Agent to initialize supervision loops and monitoring thresholds upon slice activation.
Multidomain metric streams are continuously collected and forwarded across the \textit{OBS-I1} interface to the control plane.
The \ac{SLA}-Agent evaluates these telemetry indicators to detect performance drift, resource contention, and \ac{SLA} compliance violations.
When service degradation occurs, assurance routines dispatch diagnostic events to observability endpoints to trigger corrective actions.

%% file: sections/7-Methodology.tex
% =========================
\section{Evaluation Methodology}
\label{sec:evaluation_methodology}
% =========================

This section presents the methodology adopted to evaluate the proposed \textit{TriSLA} architecture.
The experimental study leverages the operational prototype described in Section~\ref{sec:prototype_implementation}, deployed as a cloud-native application on a Kubernetes-based environment.
Under a fixed software, telemetry, and policy configuration, the evaluation covers the complete service lifecycle.
This evaluation encompasses ontology-driven semantic \ac{SLA} interpretation, \ac{ML}-based feasibility inference, \ac{XAI} attribution, preventive admission control, runtime closed-loop assurance, and end-to-end admission processing.
Throughout all experiments, the deployment configuration remained unchanged, ensuring that the observed behavior reflects controlled variations in multidomain operating conditions rather than modifications to the underlying implementation.

All reported evaluations rely on a single consolidated experimental dataset generated from telemetry, execution, admission, and runtime records collected under controlled multidomain operating conditions throughout the service lifecycle.
This dataset integrates semantic processing records, multidomain telemetry snapshots, admission decisions, runtime assurance events, and end-to-end workflow measurements into a unified experimental foundation.
Objective-specific subsets extracted from this dataset are used to evaluate the different stages of the \textit{TriSLA} architecture, ensuring methodological consistency, reproducibility, and comparability across all reported results.
Consequently, the same experimental foundation supports the analyzes of semantic processing latency and robustness, \ac{ML} benchmarking, preventive admission control, runtime assurance, and end-to-end service lifecycle performance presented in Section~\ref{sec:results}.

% =========================================================
\subsection{Multidomain Emulation Testbed}
\label{subsec:experimental_environment}
% =========================================================

To evaluate \textit{TriSLA} under controlled operating conditions without requiring physical radio hardware or physical switches, the experimental testbed extends the \ac{NASP} cloud-simulated platform~\cite{Grings2026NASP}.
The testbed is instantiated across five isolated Kubernetes namespaces on a multi-node cluster.
This deployment separates the \texttt{trisla} control plane and \ac{SLA} services from the emulated network domains and observability services, as summarized in Table~\ref{tab:trisla_namespaces}.
This architectural separation isolates domain workloads while providing reproducible conditions for cross-domain slice provisioning, stress injection, and telemetry extraction.

\begin{table}[!h]
\centering
\caption{Operational namespaces in the \textit{TriSLA} experimental testbed.}
\label{tab:trisla_namespaces}
\renewcommand{\arraystretch}{1.15}
\begin{tabular}{p{0.28\linewidth} p{0.62\linewidth}}
\toprule
\textbf{Namespace} & \textbf{Purpose} \\
\midrule
\rowcolor{LightGray}
\texttt{trisla} & Control plane and \ac{SLA} services \\
\rowcolor{white}
\texttt{ns-core} & free5GC core network functions \\
\rowcolor{LightGray}
\texttt{ueransim} & \ac{RAN} emulation and \ac{UE} workloads \\
\rowcolor{white}
\texttt{nasp-transport} & Transport-domain integration \\
\rowcolor{LightGray}
\texttt{monitoring} & Prometheus, Grafana, OpenTelemetry, and Jaeger \\
\bottomrule
\end{tabular}
\end{table}

The \texttt{ns-core} namespace instantiates the 5G core network functions utilizing the open-source free5GC~\cite{Free5GC2023Docs} v3.1.1 platform.
This deployment incorporates the \ac{AMF}, \ac{SMF}, \ac{UPF}, \ac{NRF}, \ac{PCF}, and \ac{NSSF}, supported by a MongoDB instance for subscriber and session data management.
These core functions manage the \ac{PDU} session establishments, enforce slice-level policy rules, and handle user-plane packet forwarding across isolated network slices.
The containerized core functions expose standard \ac{SBI} endpoints consumed by the \textit{TriSLA} execution layer during automated slice provisioning.

The \texttt{ueransim} namespace runs UERANSIM v4.2.1 workloads to provide high-fidelity emulation of \ac{RAN} components, including containerized \ac{gNB} and \ac{UE} instances.
The simulated \ac{gNB} nodes establish standard \ac{NGAP} and \ac{GTP-U} tunnels with the \ac{AMF} and \ac{UPF} instances hosted in the core domain.
The emulated \acp{UE} dynamically generates traffic streams that match different service profiles, such as \ac{eMBB} and \ac{URLLC}, allowing realistic load injection during validation campaigns.
This setup facilitates the injection of controlled radio-domain stress and allows access latency and throughput dynamics to be measured under fluctuating slice demands.

The \texttt{nasp-transport} namespace encapsulates the \ac{TN} integration layer by deploying Mininet~\cite{Lantz2010Mininet} and ONOS \ac{SDN} controller instances.
This namespace extends the \ac{NASP} platform~\cite{Grings2026NASP} to configure programmable OpenFlow switches and establish isolated transport paths interconnecting the \ac{RAN} and core domains.
Moreover, dynamic bandwidth provisioning, priority queue assignment, and controlled link delay injection are enforced programmatically across emulated transport topologies.
Through this integration, \textit{TriSLA} evaluates the availability of transport-domain resources and detects bottleneck conditions on intermediate fronthaul and backhaul links.

The \texttt{monitoring} namespace provides full-stack observability by deploying Prometheus~\cite{Prometheus2023Docs}, Grafana, OpenTelemetry, and Jaeger services.
Telemetry collectors periodically aggregate multidomain metrics from the \ac{RAN}, \ac{TN}, and \ac{5GC} workloads, evaluating PromQL alerting rules and extracting end-to-end distributed traces.
These metrics feed into the \ac{SLA}-Agent layer to monitor the adherence to runtime performance and compute threshold violation indicators in real-time.
Deployment automation for all operational namespaces relies on Helm charts, YAML manifests, and Kubernetes ConfigMaps to coordinate endpoints across the testbed.

% =========================================================
\subsection{Multidomain Telemetry and Workload Generation}
\label{subsec:multidomain_telemetry}
% =========================================================

Multidomain telemetry snapshots were collected before each slice admission request and associated with the corresponding admission decisions and runtime supervision records.
Prometheus-based monitoring collected infrastructure metrics from the \ac{RAN}, \ac{TN}, and \ac{5GC} domains, including the \ac{PRB} utilization, \ac{TN} latency, packet loss, jitter, and \ac{5GC} CPU and memory utilization.
During request processing, workflow timestamps and lifecycle events were automatically captured across the semantic processing, \ac{ML} inference, admission control, and runtime assurance stages, enabling the reconstruction of the complete admission workflow.
Moreover, synthetic workloads and stress conditions were injected using \texttt{iperf3} for cross-traffic generation, Linux resource controllers for compute stress, and simulated connection bursts for access load.
In this context, the resulting telemetry and execution records were integrated into the consolidated experimental dataset used throughout the evaluation.

% =========================================================
\subsection{Experimental Scenarios}
\label{subsec:experimental_scenarios}
% =========================================================

The evaluation was conducted under eight controlled multidomain operating scenarios (C0--C7), representing nominal and stressed conditions across the \ac{RAN}, \ac{TN}, and \ac{5GC} domains, as detailed in Table~\ref{tab:experimental_scenarios}.
Controlled stress conditions were introduced across domains using domain-specific workload generators.
Specifically, \ac{RAN} stress is induced through simulated \ac{UE} connection bursts, while \ac{5GC} stress is generated via container CPU and memory workload injectors.
Moreover, \ac{TN} variations were introduced in Mininet and ONOS via synthetic \texttt{iperf3} cross-traffic generation, link capacity throttling, and induced packet delay.
\ac{TN} metrics including \ac{TN} path latency, jitter, packet loss rate, and link throughput utilization are monitored via Prometheus alongside \ac{PRB} usage and \ac{5GC} resources.
These scenarios constitute the operational basis for evaluating semantic processing latency, \ac{ML}-based feasibility inference, preventive admission behavior, and end-to-end service lifecycle performance.

\begin{table}[!h]
\centering
\caption{Multidomain operating scenarios used during the experimental evaluation.}
\label{tab:experimental_scenarios}
\renewcommand{\arraystretch}{1.15}
\begin{tabular}{p{0.14\linewidth} p{0.20\linewidth} p{0.20\linewidth} p{0.20\linewidth}}
\toprule
\textbf{Scenario} & \textbf{RAN} & \textbf{TN} & \textbf{5GC} \\
\midrule
\rowcolor{LightGray}
C0 & Normal & Normal & Normal \\
\rowcolor{white}
C1 & Stress & Normal & Normal \\
\rowcolor{LightGray}
C2 & Normal & Stress & Normal \\
\rowcolor{white}
C3 & Normal & Normal & Stress \\
\rowcolor{LightGray}
C4 & Stress & Stress & Normal \\
\rowcolor{white}
C5 & Normal & Stress & Stress \\
\rowcolor{LightGray}
C6 & Stress & Normal & Stress \\
\rowcolor{white}
C7 & Stress & Stress & Stress \\
\bottomrule
\end{tabular}
\end{table}

The comparative evaluation of \ac{ML} models and semantic robustness uses telemetry observations collected across all eight operating scenarios (C0--C7).
Specifically, compound multidomain stress scenarios (C4--C7) introduce concurrent degradations across two or all three domains, testing semantic mapping resilience and the boundaries of predictive feasibility under joint resource pressure.
In contrast, runtime assurance evaluation focuses on nominal and single-domain deviation scenarios (C0--C3) to assess closed-loop remediation under controlled conditions.
This partitioning enables targeted evaluation of recovery and revalidation dynamics without confounding interference from concurrent multidomain degradations.

% =========================================================
\subsection{Evaluation Metrics}
\label{subsec:evaluation_metrics}
% =========================================================

To evaluate the proposed \textit{TriSLA} architecture, multiple performance dimensions are assessed throughout the \ac{SLA} admission workflow.
First, the semantic processing stage is evaluated in terms of execution latency, slice classification accuracy, canonical template mapping, and \ac{SLA} attribute consistency across diverse natural-language inputs.
This evaluation verifies the robustness of domain ontologies and fallback mechanisms under varying linguistic complexity.
Second, \ac{ML}-based feasibility inference is characterized using prediction accuracy, precision, recall, F1-score, inference latency, and \ac{XAI} attribution latency.
Third, preventive admission control is assessed using admission decision breakdowns across outcome categories and \ac{SLA} satisfaction rates relative to reactive and static threshold baselines.
Fourth, runtime closed-loop assurance is evaluated by measuring anomaly-detection latency, recovery-policy execution time, and post-recovery revalidation latency.
Finally, end-to-end service performance is evaluated through cumulative admission latency and its decomposition into five benchmarking macro steps (M01--M05).
These macro steps measure semantic intake (M01), predictive decision (M02), multidomain provisioning (M03), observability binding (M04), and response finalization (M05) across the admission stages (Phases~1--4) established in Section~\ref{subsec:e2e_workflow}.

%% file: sections/8-Results.tex
% =========================================================
\section{Experimental Results}
\label{sec:results}
% =========================================================

We evaluate the complete \textit{TriSLA} lifecycle by first measuring the semantic processing latency and ontology robustness of incoming \ac{SLA} requests in Section~\ref{subsec:results_semantic_processing}.
Using these parsed profiles, we benchmark the multidomain feasibility inference accuracy of \ac{ML} models in Section~\ref{subsec:results_ml_feasibility_inference}.
We then characterize the computational latency of predictive inference and \ac{XAI} feature attribution in Section~\ref{subsec:results_feasibility_inference_latency}.
Next, we evaluate the system-level performance of preventive admission control in protecting active slice resources in Section~\ref{subsec:results_preventive_admission}.
Moreover, we investigate the closed-loop recovery capabilities and cycle times during runtime \ac{SLA} assurance in Section~\ref{subsec:results_runtime_assurance}.
Finally, we analyze the consolidated end-to-end latency budget across all workflow stages to verify production viability in Section~\ref{subsec:results_e2e_latency}.

\subsection{Semantic SLA Processing Latency and Robustness}
\label{subsec:results_semantic_processing}

\begin{figure}[!h]
\centering
\includegraphics[width=\linewidth]{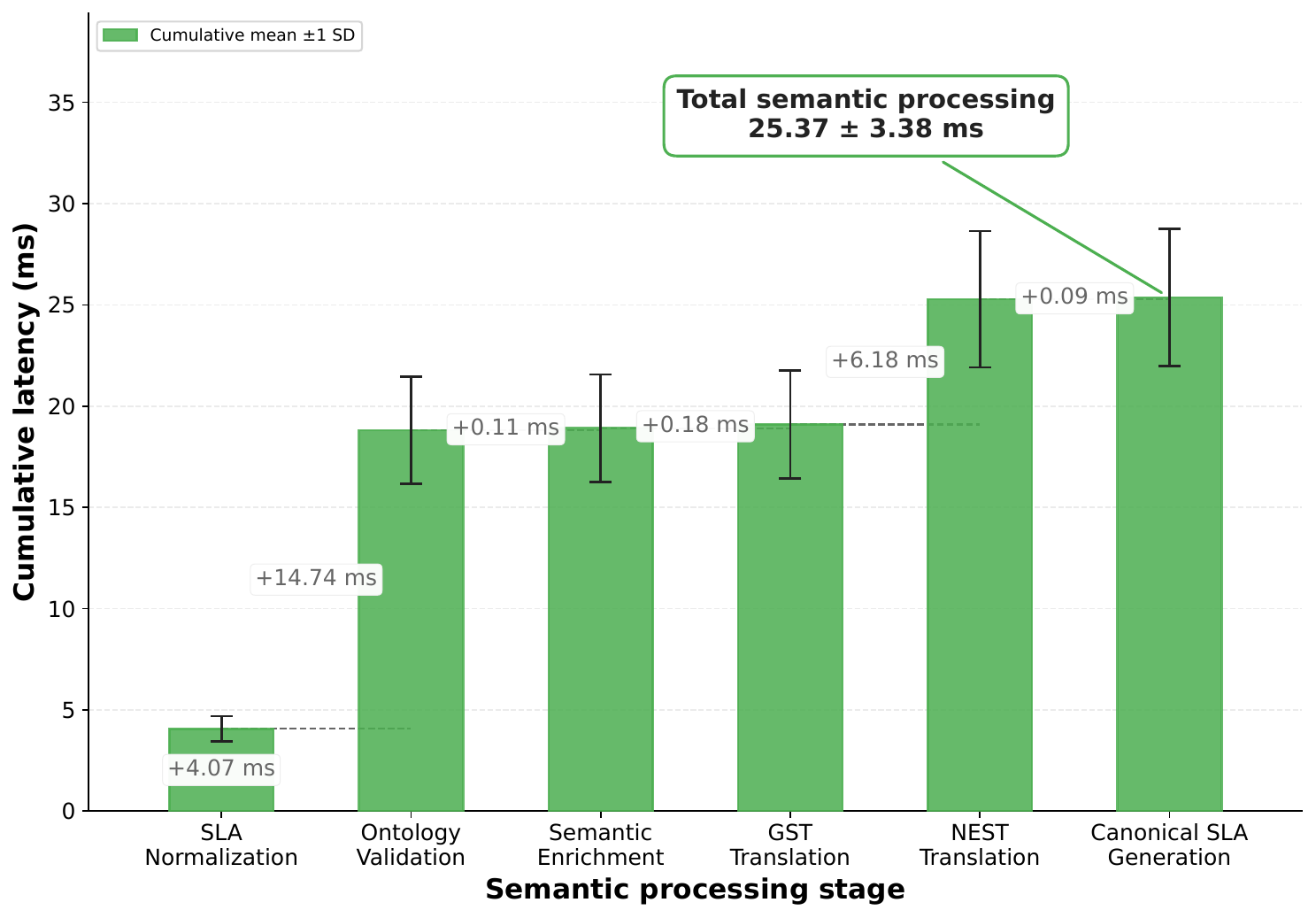}
\caption{Cumulative latency across semantic processing stages.}
\label{fig:cumulative_latency}
\end{figure}

Semantic transformation latency represents the computational duration required by the SEM-\ac{CSMF} to ingest unstructured or natural-language tenant requirements, validate them against domain ontologies, and produce standardized, machine-enforceable slice templates.
The \textit{TriSLA} pipeline completes this end-to-end semantic transformation with an average latency of $25.37 \pm 3.38$\,ms, as illustrated in Fig.~\ref{fig:cumulative_latency}.
In conventional network slicing architectures, \ac{SLA} negotiation and service template mapping rely predominantly on static catalog lookups or human-in-the-loop workflows that require minutes or hours to resolve~\cite{Li2021SLA, ETSI2022ZSM002}.
Furthermore, automated intent-based networking frameworks that employ complex description logic reasoners or iterative large-language-model parsing frequently incur processing overheads ranging from hundreds of milliseconds to multiple seconds~\cite{Leivadeas2023Intent, Njah2025AI}.
\textit{TriSLA} achieves sub-second processing by decomposing semantic interpretation into discrete sequential stages.
The workflow initiates with \ac{SLA} normalization ($4.07$\,ms) and ontology validation ($14.74$\,ms) to guarantee schema and parameter conformance.
Semantic enrichment subsequently injects domain metadata in $0.11$\,ms, followed by \ac{GST} translation ($0.18$\,ms) and \ac{NEST} template generation ($6.18$\,ms).
Finally, canonical \ac{SLA} serialization completes the pipeline in $0.09$\,ms.
The total overhead of $25.37$\,ms consumes less than $10\%$ of the overall cognitive admission target.
Because this duration remains well below standard \ac{SMO} dispatch intervals~\cite{Polese2023ORAN, 3GPP2023TS28531}, semantic reasoning executes synchronously within online slice admission without introducing operational bottlenecks.

We also evaluate the semantic processing engine across the eight scenarios (C0--C7) representing varying linguistic and operational complexities.
For slice-type service identification, the semantic classifier achieves maximum accuracy and a macro F1-score of $1.00 \pm 0.00$ across all scenarios.
The classifier correctly categorizes each request into its corresponding slice type (\ac{URLLC}, \ac{eMBB}, or \ac{mMTC}).
To characterize translation performance, we evaluate two complementary metrics: canonical \ac{SLA} mapping success and \ac{SLA} attribute consistency.
The first metric measures strict schema conformance, representing the proportion of requests in which all parameters are fully resolved to standardized \ac{GST} and \ac{NEST} templates without fallback rules.
The second metric quantifies semantic preservation, measuring the percentage of individual \ac{QoS} parameters accurately extracted from tenant intent.
This dual evaluation distinguishes between complete syntactic template compliance and parameter-level extraction of explicit service requirements.

\begin{figure}[!h]
\centering
\includegraphics[width=\linewidth]{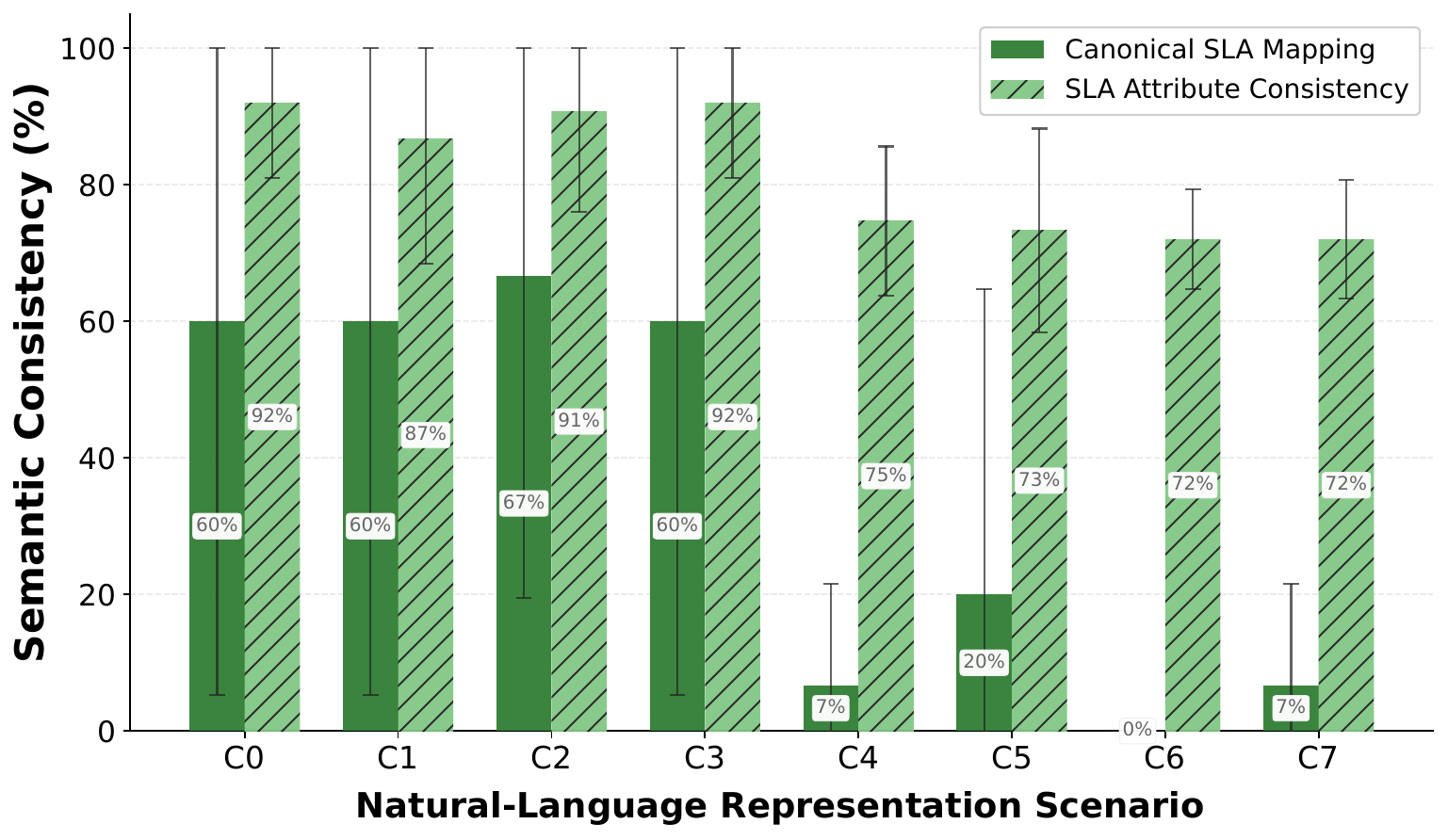}
\caption{Semantic consistency and mapping performance across natural-language representation scenarios (C0--C7).}
\label{fig:semantic_consistency_bar}
\end{figure}

Underlying \ac{SLA} attribute consistency remains high across all scenarios, ranging from $92.0\%$ in C0 and C3 to $72.0\%$ in C6 and C7, as illustrated in Fig.~\ref{fig:semantic_consistency_bar}.
This outcome confirms that the SEM-\ac{CSMF} reliably extracts individual \ac{QoS} constraints, even when tenant requests contain multi-clause natural-language specifications.
In comparison, strict canonical \ac{SLA} mapping achieves success rates between $60.0\%$ and $66.7\%$ under nominal and single-domain conditions (C0--C3).
Under compound multidomain stress, canonical mapping decreases to $6.67\%$ in C4 (\ac{RAN}+\ac{TN}) and C7 (\ac{RAN}+\ac{TN}+\ac{5GC}).
Similarly, canonical mapping reaches $20.0\%$ in C5 (\ac{TN}+\ac{5GC}) and $0.0\%$ in C6 (\ac{RAN}+\ac{5GC}).
Because canonical mapping requires an all-or-nothing match across all standard template attributes, an ambiguous or omitted descriptor in a single technical domain prevents direct schema completion.

In the compound scenario (C7), mapping performance varies across slice types.
Specifically, \ac{URLLC} requests maintain a canonical mapping success of $20.0\%$ and an attribute consistency of $84.0\%$, benefiting from concise latency bounds.
Conversely, \ac{eMBB} and \ac{mMTC} requests exhibit $0.0\%$ direct canonical mapping and lower attribute consistency ($68.0\%$ and $64.0\%$, respectively) due to high combinatorial parameter complexity.
To overcome schema incompleteness without rejecting valid tenant intents, the SEM-\ac{CSMF} leverages domain ontologies to execute deterministic fallback rules.
These rules inject standardized default parameters for unmentioned non-critical fields while strictly preserving all extracted \ac{QoS} constraints.
This ontological reasoning generates complete, machine-enforceable \ac{NEST} profiles, ensuring automated operational continuity for downstream admission control.
%Having validated the latency and semantic consistency of the intake stage, we next evaluate \ac{ML}-based feasibility inference.

\subsection{ML-Based Feasibility Inference}
\label{subsec:results_ml_feasibility_inference}

To verify the model-agnostic ML-\ac{NSMF} inference interface, we evaluate the default Random Forest classifier~\cite{Breiman2001RandomForests} against four baselines: XGBoost~\cite{Chen2016XGBoost}, LightGBM, \ac{LSTM}, and an \ac{MLP}.
The classification models were trained and validated using telemetry records collected across all eight operating scenarios (C0--C7) to ensure they effectively learn isolated and combined multidomain stress conditions.
Table~\ref{tab:ml_feasibility_metrics} summarizes the accuracy, precision, recall, and F1-score with their $95\%$ confidence intervals.
The default Random Forest model shows high classification accuracy, yielding $98.68\% \pm 0.48\%$ accuracy and $98.47\% \pm 0.72\%$ precision.
The XGBoost model achieves the highest performance, yielding $99.51\% \pm 0.33\%$ accuracy and $99.71\% \pm 0.20\%$ precision, which minimizes false admission decisions.
LightGBM also exhibits strong predictive capabilities, with an accuracy of $99.17\% \pm 0.56\%$.
Conversely, the \ac{LSTM} and \ac{MLP} show lower performance, with the \ac{MLP} yielding the lowest accuracy ($71.81\% \pm 2.00\%$) and F1-score ($50.50\% \pm 2.44\%$).

\begin{table}[!h]
\centering
\caption{Summary of Feasibility Classification Metrics across Model Architectures (Mean $\pm$ 95\% CI).}
\label{tab:ml_feasibility_metrics}
\renewcommand{\arraystretch}{1.15}
\resizebox{\columnwidth}{!}{%
\begin{tabular}{lcccc}
\hline
\textbf{Model} & \textbf{Accuracy (\%)} & \textbf{Precision (\%)} & \textbf{Recall (\%)} & \textbf{F1-Score (\%)} \\ \hline
\rowcolor{LightGray!100}
ML-NSMF (R. Forest) & $98.68 \pm 0.48$ & $98.47 \pm 0.72$ & $97.32 \pm 0.95$ & $97.79 \pm 0.80$ \\
\rowcolor{white}
XGBoost             & $99.51 \pm 0.33$ & $99.71 \pm 0.20$ & $99.48 \pm 0.36$ & $99.59 \pm 0.29$ \\
\rowcolor{LightGray!100}
LightGBM            & $99.17 \pm 0.56$ & $98.95 \pm 0.83$ & $99.03 \pm 0.65$ & $98.92 \pm 0.79$ \\
\rowcolor{white}
LSTM                & $92.78 \pm 1.35$ & $92.26 \pm 1.66$ & $89.78 \pm 2.22$ & $90.62 \pm 1.95$ \\
\rowcolor{LightGray!100}
MLP                 & $71.81 \pm 2.00$ & $55.06 \pm 2.43$ & $52.73 \pm 2.31$ & $50.50 \pm 2.44$ \\ \hline
\end{tabular}%
}
\end{table}

These results indicate that ensemble trees and neural networks can effectively learn the complex boundary constraints of multi-domain infrastructures.
However, the performance of \ac{LSTM} and \ac{MLP} indicates that sequential modeling and basic multi-layer perceptrons are less suited for immediate feasibility decisions.
Such decisions rely on current telemetry states rather than temporal dependencies.
The Random Forest model remains the default classifier in our prototype due to its balanced predictive accuracy, high precision, and resilience to fluctuations in telemetry.
Crucially, its ensemble tree structure enables polynomial-time \ac{XAI} algorithms via TreeSHAP~\cite{Lundberg2020TreeSHAP}.
This capability delivers low-latency \ac{XAI} feature attributions that explain contributions to multidomain bottlenecks without delaying online admission decisions.
%Next, we analyze the execution latency of this feasibility inference and explainability pipeline.

\subsection{Computational Latency of Feasibility Inference and XAI Attribution}
\label{subsec:results_feasibility_inference_latency}

\begin{figure}[!h]
\centering
\includegraphics[width=\linewidth]{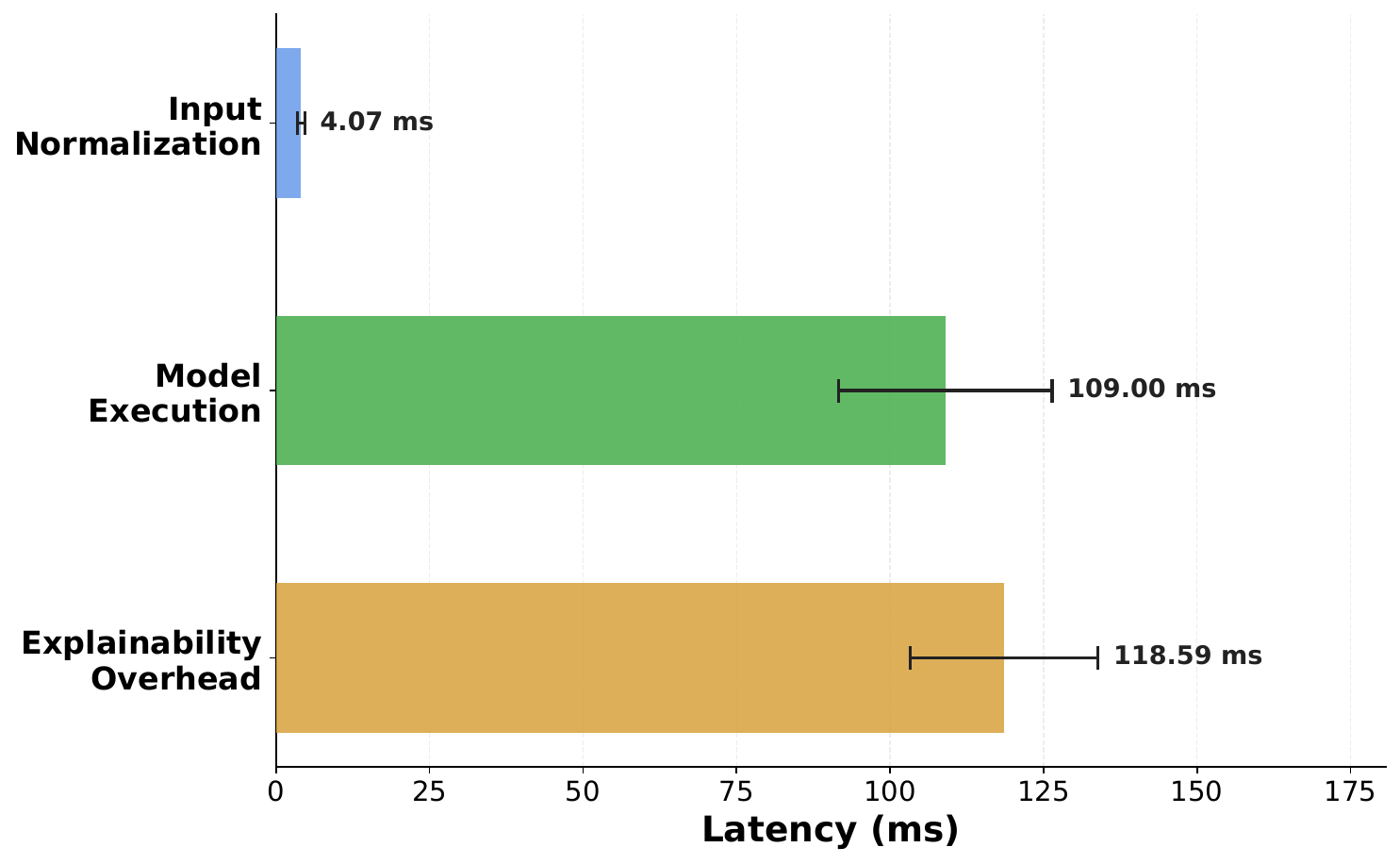}
\caption{Processing latency breakdown for feasibility inference and XAI feature attribution within the ML-NSMF.}
\label{fig:ml_inference_latency}
\end{figure}

To ensure transparent and auditable decisions, the ML-\ac{NSMF} correlates service requirements with live multidomain telemetry, executes feasibility inference, and generates \ac{XAI} feature attributions using \ac{SHAP}~\cite{Lundberg2017SHAP, Lundberg2020TreeSHAP}.
The complete predictive admission and \ac{XAI} pipeline requires a cumulative processing time of $231.66$\,ms, as illustrated in Fig.~\ref{fig:ml_inference_latency}.
The workflow begins with telemetry feature scaling and input normalization ($4.07$\,ms), followed by predictive model execution ($109.00$\,ms) to evaluate slice feasibility.
Computing domain-level \ac{XAI} attributions via TreeSHAP requires $118.59$\,ms, delivering granular root-cause transparency within sub-second timescales.
These \ac{XAI} attributions reveal the exact feature contributions to \ac{RAN} \ac{PRB} usage, \ac{TN} packet jitter, and \ac{5GC} compute load, identifying the underlying drivers of admission rejections or renegotiations.
This execution duration is negligible compared to carrier-grade network slice provisioning and orchestration cycles, which typically require several seconds to minutes~\cite{3GPP2023TS28531, Shen2020Orchestration}.
While conventional \ac{AI}-driven slice admission frameworks operate as opaque black boxes without interpretability~\cite{Bega2021Admission, Xu2020AIAdmission}, \textit{TriSLA} embeds \ac{XAI} directly into the cognitive admission plane.
Furthermore, \ac{SHAP} \ac{XAI} attribution executes asynchronously in a background worker, enabling instantaneous admission responses in $113.07$\,ms.
Concurrently, the system logs structured \ac{XAI} metadata for operational compliance auditing, policy verification, and automated root-cause diagnostics~\cite{Brik2024XAI, Sun2025XAI}.

\subsection{Preventive Admission Control Performance}
\label{subsec:results_preventive_admission}

\begin{figure}[!h]
\centering
\includegraphics[width=\linewidth]{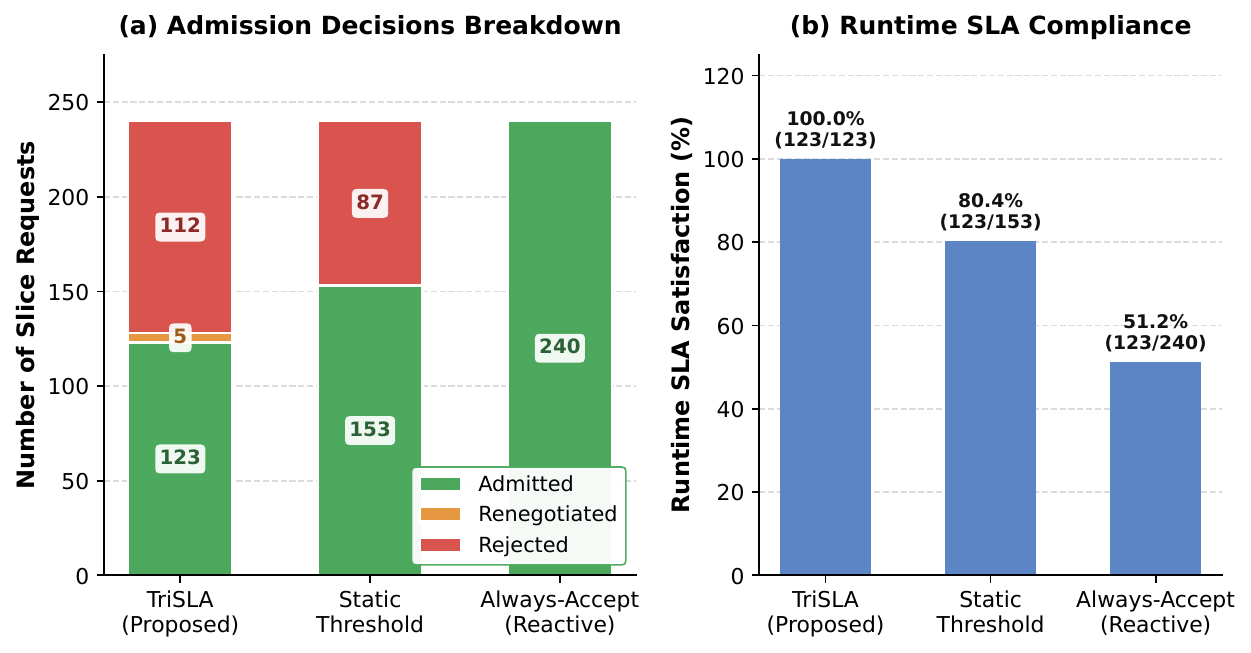}
\caption{Preventive admission control performance comparing \textit{TriSLA} against static and reactive baselines under a 240-request workload: (a) admission decisions breakdown across decision categories (admitted, renegotiated, rejected) and (b) resulting runtime SLA satisfaction rate among admitted slices.}
\label{fig:preventive_admission_performance}
\end{figure}

We evaluate the preventive admission control of \textit{TriSLA} against two comparative baseline strategies under an identical evaluation workload of 240 multidomain slice requests.
The first strategy is a static threshold baseline, which evaluates incoming slice requests against fixed, independent per-domain utilization thresholds without cross-domain correlation or profile renegotiation capabilities.
The second strategy is a reactive always-accept baseline, which unconditionally admits all arriving requests without pre-admission feasibility checks, delegating all mitigation to runtime closed-loop remediation.
Fig.~\ref{fig:preventive_admission_performance} demonstrates the two-phase cause-and-effect relationship between admission-time decision filtering (Fig.~\ref{fig:preventive_admission_performance}a) and resulting runtime \ac{SLA} compliance (Fig.~\ref{fig:preventive_admission_performance}b).

\textit{TriSLA} evaluates multidomain feasibility before resource allocation, categorizing incoming requests into three distinct decision outcomes, as depicted in Fig.~\ref{fig:preventive_admission_performance}a.
Under \textit{TriSLA}, 123 requests ($51.25\%$) are directly admitted (green segment), and five borderline requests ($2.08\%$) are dynamically renegotiated to compliant \ac{QoS} profiles (orange segment).
Additionally, 112 infeasible requests ($46.67\%$) are preventively rejected (red segment) to protect infrastructure capacity.
Together, direct admission and dynamic profile renegotiation achieve an effective acceptance of 128 slices ($53.33\%$) out of 240 requests.
However, the static threshold baseline evaluates resource limits against fixed per-domain thresholds without cross-domain feasibility prediction, admitting 153 requests ($63.75\%$) and rejecting 87 requests ($36.25\%$) without renegotiation support.
Moreover, the reactive always-accept baseline indiscriminately admits all 240 requests ($100.0\%$) without evaluating resource constraints.

A direct cause-and-effect relationship exists between the admission decisions in Fig.~\ref{fig:preventive_admission_performance}a and the runtime \ac{SLA} compliance in Fig.~\ref{fig:preventive_admission_performance}b.
\textit{TriSLA} achieves a $100.0\%$ \ac{SLA} satisfaction rate for all 123 directly admitted slices (Fig.~\ref{fig:preventive_admission_performance}b), demonstrating that predictive filtering and dynamic profile renegotiation successfully shield active slices from multidomain resource contention.
The static threshold baseline improves performance over the reactive approach but still suffers from a reduced satisfaction rate of $80.4\%$ due to false admissions during cross-domain resource contention spikes.
Specifically, only 123 of the 153 admitted slices maintain compliance under static thresholds.
Because the reactive baseline unconditionally admits the 240 requests, it triggers severe multidomain resource overprovisioning.
Consequently, it results in a degraded runtime \ac{SLA} satisfaction rate of only $51.2\%$ (123 of 240 admitted slices meet their \acp{SLA}).
While preventive admission protects active slices from initial overload, continuous closed-loop runtime assurance is necessary during service execution.

\subsection{Closed-Loop SLA Supervision and Runtime Assurance}
\label{subsec:results_runtime_assurance}

To evaluate the responsiveness and remediation efficacy of runtime assurance, \textit{TriSLA} is benchmarked across 24 execution runs under controlled telemetry deviation scenarios (C0--C3).
As established in Section~\ref{subsec:experimental_scenarios}, runtime assurance focuses on nominal conditions (C0) and single-domain stress scenarios (C1--C3) to isolate per-domain recovery dynamics without compounding cross-domain interference.
Each scenario is evaluated under two distinct operating modes: active Closed-Loop remediation and passive Monitor-Only supervision.
We analyze the Closed-Loop mode as a complete autonomic lifecycle by detecting violations and executing domain-specific recovery policies.
Moreover, the Monitor-Only mode evaluates an empirical baseline to isolate telemetry collection and observability overheads without triggering corrective actuations.

\begin{figure}[!h]
\centering
\includegraphics[width=\linewidth]{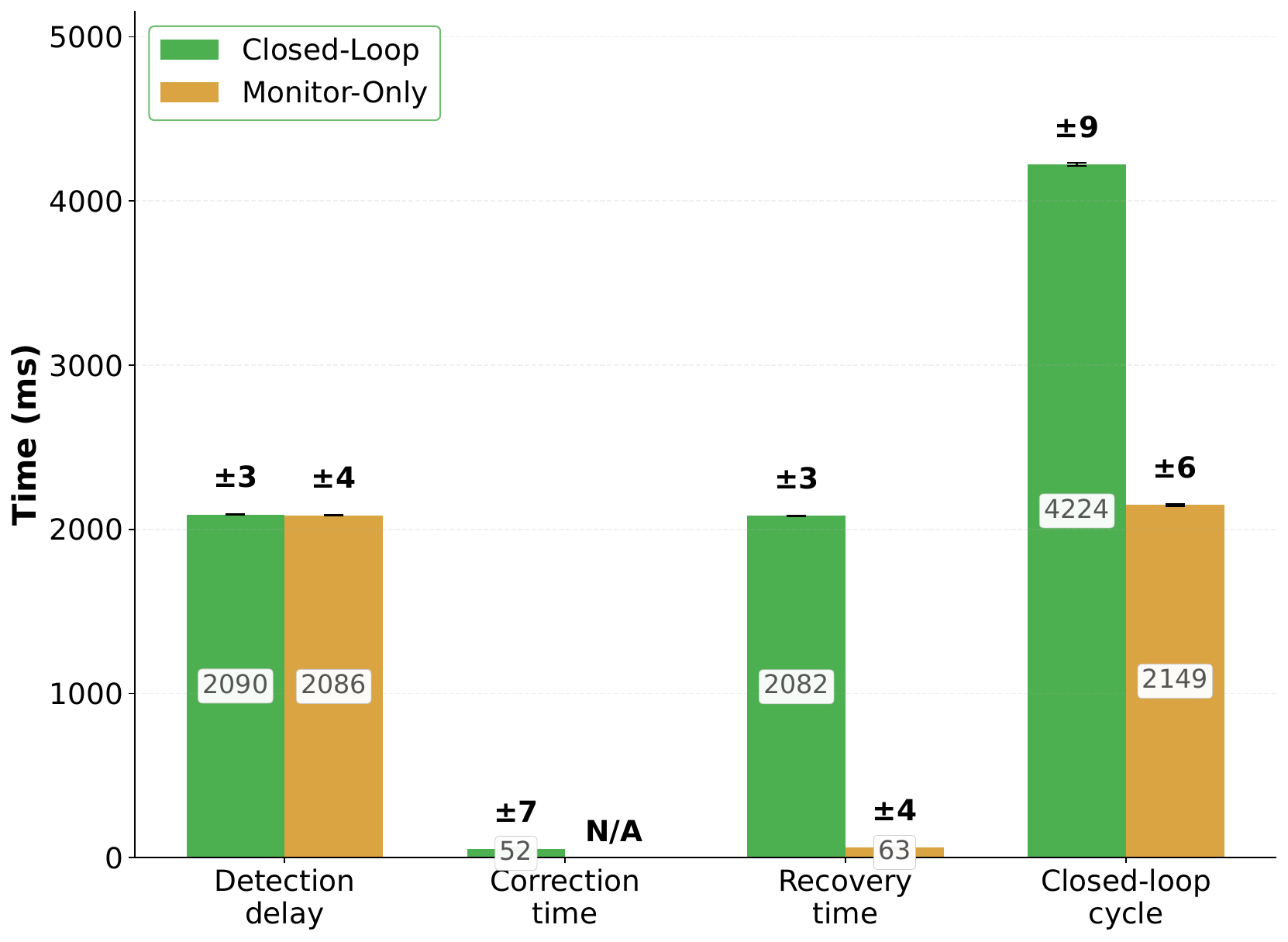}
\caption{Breakdown of runtime assurance cycle times across four operational stages comparing Closed-Loop and Monitor-Only operating modes.}
\label{fig:runtime_assurance}
\end{figure}

Fig.~\ref{fig:runtime_assurance} presents four comparative metric groups, each contrasting the Closed-Loop (green bars) and Monitor-Only (orange bars) operating modes.
These groups evaluate the sequential lifecycle stages: (i) Detection delay, (ii) Correction time, (iii) Recovery time, and (iv) Closed-loop cycle.
At the management and orchestration plane, a detection window of approximately 2\,s is well aligned with standard telemetry collection intervals~\cite{ETSI2022ZSM002, Coronado2022SLA}.
This aggregation window filters out transient traffic bursts and high-frequency stochastic jitter, preventing control-loop instability and actuator flapping.
While fast data-plane adaptations operate on millisecond timescales, a multi-second lifecycle budget represents near-real-time responsiveness for cross-domain slice reconfiguration.

Analyzing the metric groups from left to right in Fig.~\ref{fig:runtime_assurance} reveals the exact timing breakdown of each operational stage.
In the first group (Detection delay), detection latency remains statistically equivalent between modes ($2090 \pm 3$\,ms for Closed-Loop and $2086 \pm 4$\,ms for Monitor-Only).
This equivalence confirms that telemetry ingestion and threshold rule evaluation operate independently of active remediation loops.
In the second group (Correction time), the Closed-Loop mode executes the recovery policy in $52 \pm 7$\,ms, whereas the Monitor-Only mode shows no latency ($\text{N/A}$) due to the actuators being disabled.
In the third group (Recovery time), the Closed-Loop mode requires a $2082 \pm 3$\,ms revalidation window to confirm \ac{SLA} metric stabilization, whereas the Monitor-Only baseline records a $63 \pm 4$\,ms observability logging overhead.
In the fourth group (Closed-loop cycle), total execution reaches $4224 \pm 9$\,ms for Closed-Loop versus $2149 \pm 6$\,ms for Monitor-Only, with active policy execution consuming only $1.2\%$ ($52$\,ms) of the total duration.

The closed-loop cycle latency remains consistent across all evaluated single-domain stress scenarios.
Specifically, total cycle duration measures $4216$\,ms under \ac{RAN} stress (C1), $4223$\,ms under \ac{TN} stress (C2), and $4241$\,ms under \ac{5GC} stress (C3).
Under the nominal scenario (C0), the \ac{SLA}-Agent observed stable telemetry profiles with zero false-positive anomaly detections.
Under transport network stress (C2), the \ac{SLA}-Agent detected packet delay and link congestion, triggering ONOS flow rerouting to restore \ac{TN} \ac{SLA} compliance.
Across all 24 experimental evaluations, all 12 detected anomalies under Closed-Loop mode were successfully restored to compliant service operation.

\subsection{Detailed End-to-End Admission Latency Analysis}
\label{subsec:results_e2e_latency}

\begin{figure}[!h]
\centering
\includegraphics[width=\linewidth]{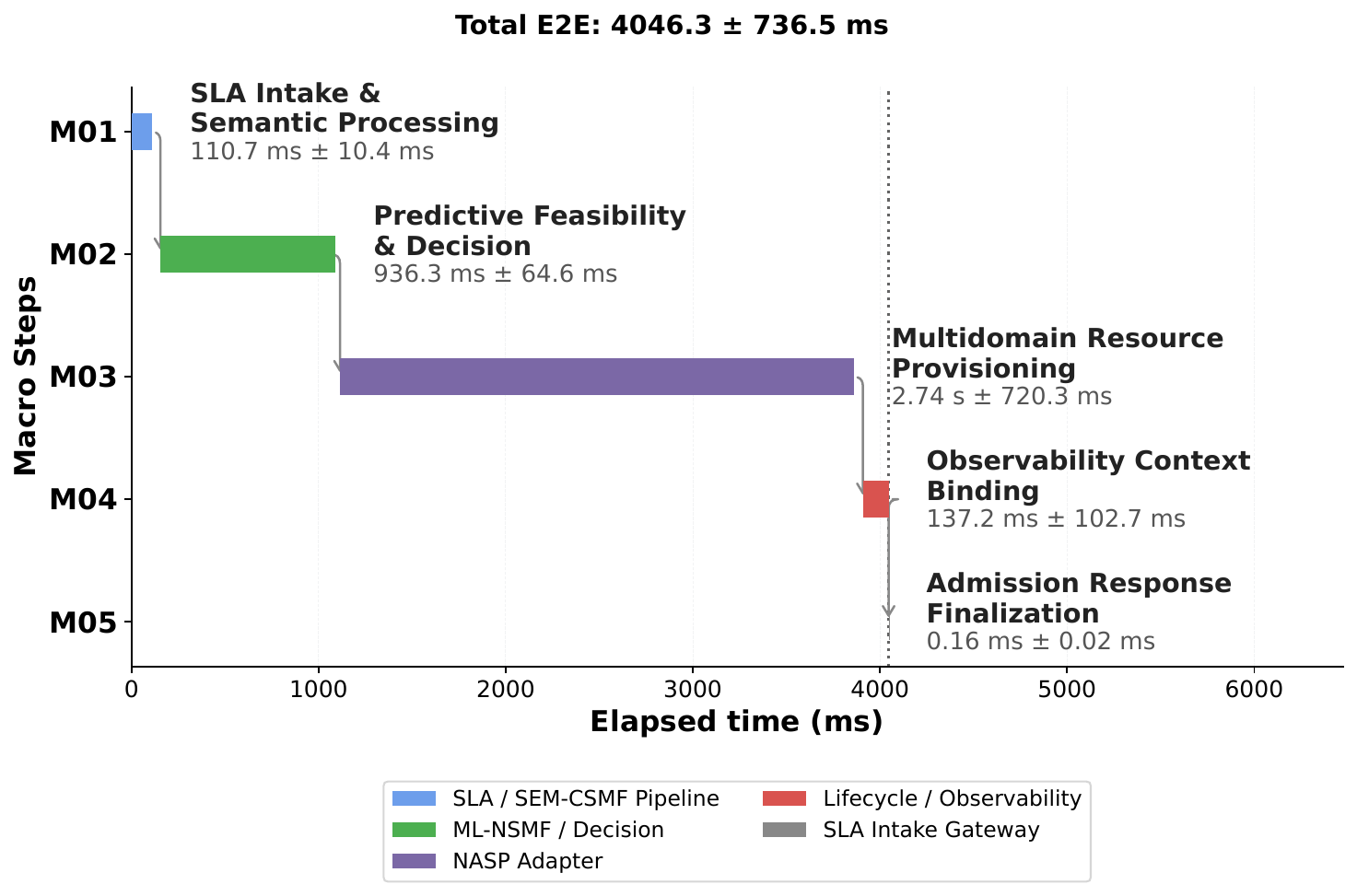}
\caption{End-to-end admission latency waterfall breakdown across sequential macro steps (M01--M05) and architectural subsystems, highlighting inter-stage communication gaps and total execution milestones.}
\label{fig:e2e_latency_breakdown}
\end{figure}

The end-to-end admission pipeline integrates semantic intake, predictive decision arbitration, multidomain provisioning, and observability context binding (Phases~1--4 in Section~\ref{subsec:e2e_workflow}).
We benchmark this integrated workflow across five sequential macro steps (M01--M05), as illustrated in the admission latency waterfall in Fig.~\ref{fig:e2e_latency_breakdown}.
The complete onboarding sequence achieves a mean \ac{E2E} latency milestone of $4046.3 \pm 736.5$\,ms.
This profile confirms that cognitive admission decisions execute in near-real-time without impeding production slice lifecycles.

The admission sequence begins with \ac{SLA} Intake and Semantic Processing (M01, $110.7 \pm 10.4$\,ms), combining intent ingestion, validation, and profiling in the SEM-\ac{CSMF}.
Following an inter-stage message serialization gap, the ML-\ac{NSMF} executes Predictive Feasibility and Decision (M02, $936.3 \pm 64.6$\,ms) to evaluate multidomain resource availability and trigger \ac{XAI} attribution.
Upon affirmative admission and inter-process dispatch, the \ac{NASP} Adapter executes Multidomain Resource Provisioning (M03, $2.74$\,s $\pm 720.3$\,ms).
Following container coordination, the \ac{SLA}-Agent completes Observability Context Binding (M04, $137.2 \pm 102.7$\,ms) to synchronize telemetry identifiers.
Finally, the \ac{SLA} Gateway executes Admission Response Finalization (M05, $0.16 \pm 0.02$\,ms) to return deployment confirmation to the tenant.

The total latency budget is predominantly governed by multidomain resource provisioning in the \ac{NASP} Adapter (M03, $2.74$\,s $\pm 720.3$\,ms).
In contrast, cognitive admission intelligence (M01 and M02) and observability context binding (M04) execute in sub-second durations.
Consequently, ontology-driven semantic reasoning, predictive inference, and \ac{XAI} attribution introduce negligible latency overhead relative to underlying infrastructure provisioning cycles.

%% file: sections/9-Conclusion.tex
% =========================================================
\section{Conclusion}
\label{sec:conclusion}
% =========================================================
This article presented \textit{TriSLA}, a preventive and closed-loop \ac{SLA}-aware architecture designed to support service admission and runtime assurance in multidomain 5G environments.
The \textit{TriSLA} architecture overcomes conventional reactive \ac{SLA} limitations by unifying ontology-driven intent translation, predictive feasibility inference, and \ac{XAI} decision attribution.
This intelligence plane operates alongside continuous runtime supervision within an integrated operational workflow.
By correlating service requirements with multidomain infrastructure conditions before service admission, \textit{TriSLA} enables preventive admission decisions while maintaining consistency throughout the service lifecycle.
In this context, the architecture was validated using a fully operational prototype deployed in a Kubernetes-based, multidomain environment integrating the \ac{RAN}, \ac{TN}, and \ac{5GC} domains.
Experimental evaluation on an operational dataset confirmed that \textit{TriSLA} delivers high feasibility prediction accuracy, transparent \ac{XAI} attributions, preventive admission guarantees, and sub-second cognitive processing.

Beyond individual functional evaluation, the results demonstrate that the semantic, predictive, and runtime assurance mechanisms of \textit{TriSLA} operate as an integrated architecture across the entire service lifecycle.
The combination of \ac{XAI} feature attribution and runtime telemetry correlation enhances the transparency and trustworthiness of admission decisions, while closed-loop control supports continuous compliance verification after service deployment.
Together, these capabilities provide a practical foundation for multidomain \ac{SLA} management in scenarios where trust, accountability, and continuous compliance are essential operational requirements.
Although validated in an operational testbed, future work will extend evaluation to larger multidomain deployments, higher request rates, and more heterogeneous network topologies.
Additional research will investigate adaptive policy refinement, online model retraining under concept drift, and reinforcement learning strategies to further optimize closed-loop control in dynamic 5G environments.

%% file: sections/9-Acknowledgment.tex
% =========================================================
\section*{Acknowledgment}
\label{sec:acknowledgment}
% =========================================================
This work was partially supported by CNPq Grants Nos.\ 405111/2021-5 and 130555/2019-3, and by CAPES, Finance Code 001, Brazil; additional support was provided by RNP and MCTIC under Grant No.\ 01245.010604/2020-14 as part of the 6G Brasil and OpenRAN@Brasil projects, and by MCTIC/CGI.br/FAPESP through Project SAMURAI (Grant No.\ 2020/05127-2) and Project PORVIR-5G (Grants No.\ 2020/05182-3 and 2025/01970-0); it has also been funded in part by the projects XGM-AFCCT-2024-5-1-1 and XGM-AFCCT-2026-5-1-1 supported by xGMobile, EMBRAPII, Inatel Competence Center on 5G and B5G Networks, with financial resources from the PPI IoT/Manufatura 4.0/the MCTI grant number 052/2023, signed with EMBRAPII; and finally, this work also received support from the Commonwealth Cyber Initiative (www.cyberinitiative.org).

%% file: sections/10-Biographies.tex
% =========================================================
% AUTHOR BIOGRAPHIES
% =========================================================

\begin{IEEEbiography}[{\includegraphics[width=1in,height=1.25in,clip,keepaspectratio]{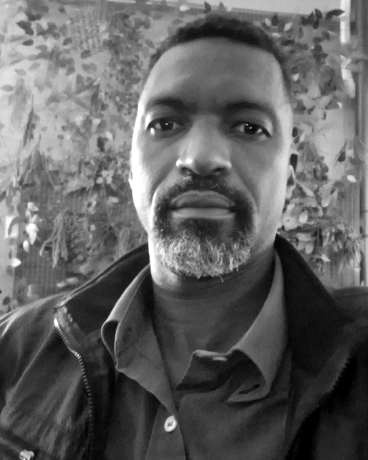}}]{Abel~J.~R.~Lisboa}
received the degree in Computer Network Technology from Sociedade Educacional Três de Maio (SETREM), Três de Maio, Rio Grande do Sul, Brazil.
He is currently pursuing the M.Sc. degree in applied computing with the Universidade do Vale do Rio dos Sinos (UNISINOS), S\~{a}o Leopoldo, Brazil.
His research interests include Open RAN, multidomain network slicing orchestration, Service Level Agreement assurance, and Explainable Artificial Intelligence in 5G and 6G systems.
\end{IEEEbiography}

\begin{IEEEbiography}[{\includegraphics[width=1in,height=1.25in,clip,keepaspectratio]{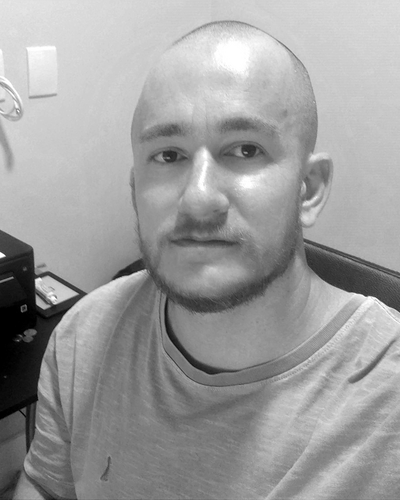}}]{Gustavo~Z.~Bruno\orcidlink{0000-0002-1424-3404}}
received the Ph.D. degree in applied computing from the Universidade do Vale do Rio dos Sinos (UNISINOS), S\~{a}o Leopoldo, Brazil.
He is currently a Postdoctoral Researcher with the Instituto Nacional de Telecomunica\c{c}\~{o}es (Inatel), Brazil, and an IT Analyst with MTI, Brazil.
His research interests include beyond 5G and 6G architectures, Open RAN, and autonomous network slice assurance.
\end{IEEEbiography}

\begin{IEEEbiography}[{\includegraphics[width=1in,height=1.25in,clip,keepaspectratio]{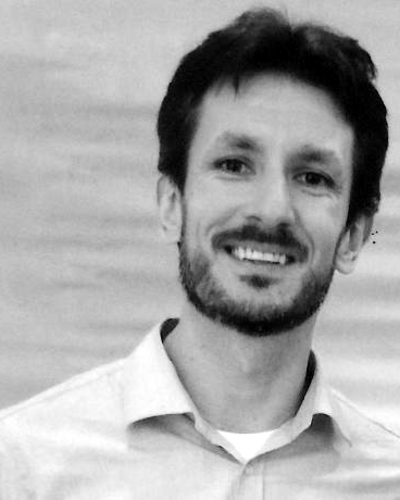}}]{Cristiano~B.~Both\orcidlink{0000-0002-9776-4888}}
(Member, IEEE) received the B.S. degree from UPF, Brazil, and the M.Sc. and Ph.D. degrees in computer science from UFRGS, Porto Alegre, Brazil.
He is a Full Professor with the Applied Computing Graduate Program at UNISINOS, Brazil, and a CNPq Research Productivity Fellow.
His research interests include SDN/NFV, wireless networks, Open RAN, and multidomain network slicing management.
\end{IEEEbiography}